\documentclass[prc,twocolumn,superscriptaddress,showpacs,preprintnumbers,amsmath,amssymb]{revtex4-1}
\usepackage[dvips]{graphicx}
\usepackage{dcolumn}
\usepackage{bm}

\begin{document}
\title{Odd-particle number random phase approximation and extensions: 
Applications to particle and hole states around $^{16}$O}
\author{Mitsuru Tohyama}
\affiliation{Kyorin University School of Medicine, Mitaka, Tokyo
  181-8611, Japan} 
\author{Peter Schuck}
\affiliation{Institut de Physique Nucl$\acute{e}$aire, IN2P3-CNRS,
Universit$\acute{e}$ Paris-Sud, F-91406 Orsay Cedex, France}
\affiliation{Laboratoire de Physique et de Mod\'elisation des Milieux Condens\'es, CNRS 
 et, Universit\'e Joseph Fourier, 25 Av. des Martyrs, BP 166, F-38042 
 Grenoble Cedex 9, France}

%\date{\today} 
\begin{abstract}
The hole-state random phase approximation (hRPA) and the particle-state random phase approximation (pRPA) for systems like odd $A$ nuclei are discussed. 
These hRPA and pRPA are formulated based on the Hartree-Fock ground state.
An extension of hRPA and pRPA
based on a correlated ground state
is given using time-dependent density-matrix theory.
Applications to the single-particle states around $^{16}$O are presented. 
It is shown that inclusion of ground-state
correlation affects appreciably the results of hRPA and pRPA. The question of 
the coupling of the center of mass motion of the core to the particle (hole) 
is also discussed.
\end{abstract}
\maketitle
\section{Introduction}
The one-particle states and one-hole states are basic excitation modes of a nucleus and other many body systems. The experimental data on 
the properties of the single-particle states have been accumulated using nuclear reactions such as 
one-nucleon transfer, pickup and knock-out reactions \cite{Diep}, and it has been found that there is a substantial depletion of the
spectral strength of the single-particle states. Theoretical studies have shown 
that the strong short-range and tensor components of the nucleon-nucleon interaction
are responsible for a part of the depletion \cite{vond} and a substantial part of the fragmentation of the single-particle strength
is due to the coupling to low-lying collective modes \cite{bortignon,bortignon1}. 
The standard approach to study the single-particle properties may be the Green's function method.
Various theoretical approaches have been proposed to implement the coupling to low-lying collective modes into
the self-energy of the Green's function: the particle-phonon coupling model \cite{Giai,bortignon,litv,colo},
the Tamm-Dancoff approximation (TDA) \cite{hrpa96} 
and the more recent Faddeev random-phase
approximation (FRPA) \cite{dick02}. In the present paper we give a 
formulation of 
the hole-state RPA (hRPA) using the equation of motion
approach (EoM) \cite{Rowe}, which has often been used to derive the standard 
RPA, and discuss some aspects of hRPA
such as the relation to the particle-state RPA (pRPA),
which have not been clarified so far in the literature \cite{hrpa73,hrpa96}.
We also present an extension 
of odd $A$ RPA (oRPA) based on a correlated ground state
obtained from the time-dependent density-matrix theory (TDDM) \cite{WC,WC1,GT}. 
The influence of the center-of-mass (c.o.m.) motion of the even core on the odd system is 
also discussed.
The paper is organized as follows: the formulation of hRPA
and its extension is given in sect. 2, some properties of the extended RPA 
are also discussed in sect.2, 
the results obtained for the single-particle states around $^{16}$O are presented 
in sect. 3, and 
sect. 4 is devoted to a discussion and conclusion section. 

\section{Formulation}
Let us consider a nucleus consisting of $A$ nucleons and assume that the 
total Hamiltonian $H$
consists of the kinetic energy term and a two-body interaction. Let us assume 
that $|0\rangle$ is the ground state of the $A$ nucleon system with $A$ even 
and with energy $E_0$ and 
$|\mu\rangle$ an exact eigenstate of the Hamiltonian 
for the $A-1$ system with an eigenvalue $E_\mu$ ($H|\mu\rangle=E_\mu|\mu\rangle$).

\subsection{Equations of motion for transition amplitudes}
In direct reaction theories such as the  
the distorted wave impulse approximation and the distorted wave Born approximation 
the differential cross section for one nucleon transfer reactions
is related to the spectral function $S_{\alpha\alpha'}(\omega)$
\begin{eqnarray}
S_{\alpha\alpha'}(\omega)=\sum_\mu\langle 0|a^+_{\alpha'}|\mu\rangle\langle \mu|a_\alpha|0\rangle
\delta(\omega+E_\mu-E_0),
\end{eqnarray}
where $a_\alpha$ and $a^+_\alpha$ are the annihilation and creation operators of a nucleon in a single-particle state $\alpha$,
respectively.
We consider the equations of motion for the transition amplitudes $x^\mu_\alpha$ and $X^\mu_{\alpha\beta:\gamma}$
from the $A$ nucleon system to the $A-1$ nucleon system. These amplitudes are defined by
\begin{eqnarray}
x^\mu_\alpha&=&\langle 0|a^+_\alpha|\mu\rangle, \\
X^\mu_{\alpha\beta:\gamma}&=&\langle 0|:a^+_\alpha a^+_\beta a_\gamma:|\mu\rangle,
\end{eqnarray}
where 
 $:~:$ implies 
\begin{eqnarray}
:a^+_\alpha a^+_\beta a_\gamma:=a^+_\alpha a^+_\beta a_\gamma-(n_{\gamma\beta}a^+_\alpha-n_{\gamma\alpha}a^+_\beta).
\end{eqnarray} 
Here, $n_{\alpha\alpha'}$ is the occupation matrix given by
\begin{eqnarray} 
n_{\alpha\alpha'}=\langle 0|a^+_{\alpha'}a_\alpha|0\rangle.
\end{eqnarray}
From the EoM relation 
\begin{eqnarray}
\langle 0|[H,a^+_\alpha]=\langle 0|a^+_\alpha(E_0-H),
\end{eqnarray}
we obtain the equation for $x^\mu_\alpha$
\begin{eqnarray}
\langle 0|[H,a^+_\alpha]|\mu\rangle&=&\omega_\mu\langle 0|a^+_\alpha|\mu \rangle=\omega_\mu x^\mu_\alpha, 
\label{eq1}
\end{eqnarray}
where $\omega_\mu=E_0-E_\mu$.
The commutator on the left-hand side of the above equation includes terms with $a^+_\alpha a^+_\beta a_\gamma$.
Therefore, $x^\mu_\alpha$ couples to $X^\mu_{\alpha\beta:\gamma}$. In a way 
analogous to that used in deriving Eq.~(\ref{eq1}), we obtain
the equation for $X^\mu_{\alpha\beta:\gamma}$
\begin{eqnarray}
\langle 0|[H,:a^+_\alpha a^+_\beta a_\gamma:]|\mu \rangle&=&\omega_\mu\langle 0|:a^+_\alpha a^+_\beta a_\gamma:|\mu\rangle
\nonumber \\
&=&\omega_\mu X^\mu_{\alpha\beta:\gamma}.
\label{eq2}
\end{eqnarray}
On the left-hand side of the above equation there appear
expectation values of the
terms consisting of three creation operators and two annihilation operators
such as $\langle 0|a^+_{\lambda_1} a^+_{\lambda_2} a^+_{\lambda_3} a_{\lambda_4} a_{\lambda_5}|\mu\rangle$,
which implies the coupling to a higher-level amplitude
$\langle 0|:a^+_{\alpha} a^+_{\beta} a^+_{\gamma} a_{\beta'} a_{\alpha'}:|\mu \rangle$.
To close the chain of the coupled equations,
we factorize these terms using  $x^\mu_\alpha$ and $X^\mu_{\alpha\beta:\gamma}$ as
\begin{eqnarray}
\langle 0|a^+_{\lambda_1} a^+_{\lambda_2} a^+_{\lambda_3} a_{\lambda_4} a_{\lambda_5}|\mu\rangle
&\approx& {\cal AS}(n_{\lambda_4\lambda_3}X^\mu_{\lambda_1\lambda_2:\lambda_5}
\nonumber \\
&+&
C_{\lambda_5\lambda_4\lambda_2\lambda_3}x^\mu_{\lambda_1}),
\end{eqnarray}
where the correlation matrix $C_{\alpha\beta\alpha'\beta'}$ is defined by
$C_{\alpha\beta\alpha'\beta'}=\langle 0|:a^+_{\alpha'}a^+_{\beta'}a_\beta a_\alpha:|0\rangle$
and ${\cal AS(~)}$ means that the terms in the parentheses are properly antisymmterized \cite{WC}. 
The obtained coupled equations are written as
\begin{eqnarray}
(\epsilon_\alpha -\omega_\mu)x^\mu_\alpha&+&\sum_{\lambda_1\lambda_2\lambda_3}\langle\lambda_1\lambda_2|v|\alpha\lambda_3\rangle
X^\mu_{\lambda_1\lambda_2:\lambda_3}=0,
\label{stddm1}
\end{eqnarray}
\begin{eqnarray}
(\epsilon_\alpha&+&\epsilon_\beta-\epsilon_\gamma-\omega_\mu )X^\mu_{\alpha\beta:\gamma}+\sum_{\lambda_1\lambda_2\lambda_3\lambda_4}
\langle\lambda_1\lambda_2|v|\lambda_3\lambda_4\rangle_A
\nonumber \\
&\times&[(\delta_{\lambda_3\alpha}((\delta_{\lambda_4\beta}-n_{\lambda_4\beta})n_{\gamma\lambda_2}-C_{\gamma\lambda_4\lambda_2\beta})
\nonumber \\
&+&\delta_{\lambda_3\beta}(n_{\lambda_4\alpha}n_{\gamma\lambda_2}+C_{\gamma\lambda_4\lambda_2\alpha})
\nonumber \\
&+&\delta_{\lambda_2\gamma}(n_{\lambda_3\alpha}n_{\lambda_4\beta}+\frac{1}{2}C_{\lambda_3\lambda_4\alpha\beta})) x^\mu_{\lambda_1}
\nonumber \\
&+&\delta_{\lambda_3\alpha}n_{\gamma\lambda_1}X^\mu_{\lambda_2\beta:\lambda_4}
+\delta_{\lambda_3\beta}n_{\gamma\lambda_2}X^\mu_{\lambda_1\alpha:\lambda_4}
\nonumber \\
&-&\delta_{\lambda_1\gamma}(n_{\lambda_4\alpha}X^\mu_{\beta\lambda_2:\lambda_3}
+n_{\lambda_3\beta}X^\mu_{\alpha\lambda_2:\lambda_4})
\nonumber \\
&+&\frac{1}{2}(\delta_{\lambda_3\alpha}\delta_{\lambda_4\beta}-\delta_{\lambda_3\alpha}n_{\lambda_4\beta}
+\delta_{\lambda_3\beta}n_{\lambda_4\alpha})X^\mu_{\lambda_1\lambda_2:\gamma}]
\nonumber \\
&=&0,
\label{stddm2}
\end{eqnarray}
where the subscript $A$ means that the corresponding matrix is antisymmetrized and the single-particle states are chosen as the eigenstates of the matrix 
\begin{eqnarray}
\langle\alpha|t|\alpha'\rangle+\sum_{\lambda\lambda'}\langle\alpha\lambda|v|\alpha'\lambda'\rangle_An_{\lambda'\lambda}.
\end{eqnarray}
Here $t$ is the kinetic energy operator.
Equations (\ref{stddm1}) and (\ref{stddm2}) are written in matrix form:
\begin{eqnarray}
\left(
\begin{array}{cc}
a&c\\
b&d
\end{array}
\right)\left(
\begin{array}{c}
x^\mu\\
X^\mu
\end{array}
\right)
=\omega_\mu
\left(
\begin{array}{c}
x^\mu\\
X^\mu
\end{array}
\right).
\label{hrpa}
\end{eqnarray}
The matrix elements of the above equation are given in Appendix A.
The normalization of the amplitudes is given by 
\begin{eqnarray}
\left(\tilde{x}^{\mu~*}~\tilde{Y}^{\mu~*}\right)
\left(
\begin{array}{c}
X^{\mu'}\\
Y^{\mu'}
\end{array}
\right)=\delta_{\mu\mu'},
\label{normhrpa}
\end{eqnarray}
where $\tilde{x}^{\mu~*}_{\alpha\alpha'}$ and $\tilde{X}^{\mu~*}_{\alpha\beta\alpha'\beta'}$ 
are the left eigenvector of Eq. (\ref{hrpa}).
The occupation matrix and the correlation matrix, which enter Eq. (\ref{hrpa}) and which 
describe the ground-state correlations in the $A$ nucleon system, 
can be determined in the framework of Time Dependent Density Matrix (TDDM) 
theory: the TDDM equations \cite{WC,GT} consist of the coupled equations 
motion for
$n_{\alpha\alpha'}$ and $C_{\alpha\beta\alpha'\beta'}$,
\begin{eqnarray}
i\hbar\dot{n}_{\alpha\alpha'}&=&\langle 0|[a^+_{\alpha'}a_\alpha,H] 0\rangle,
\label{tddm1}
\end{eqnarray}
\begin{eqnarray}
i\hbar\dot{C}_{\alpha\beta\alpha'\beta'}&=&\langle 0|[:a^+_{\alpha'}a^+_{\beta'}a_\beta a_\alpha:,H]|0\rangle.
\label{tddm2}
\end{eqnarray}
The right-hand side of Eq.~(\ref{tddm2}) contains the expectation values of 
three-body operators, which are
approximated by the products of $n_{\alpha\alpha'}$ and $C_{\alpha\beta\alpha'\beta'}$ 
to close the coupled chain
of the equations of motion. The ground state in TDDM is given as a stationary 
solution of the TDDM equations which
satisfies $\dot{n}_{\alpha\alpha'}=0$ and $\dot{C}_{\alpha\beta\alpha'\beta'}=0$. 
The stationary solution can be obtained 
using the gradient method \cite{Toh}. This method will be used in our 
numerical application given later.

In the Hartree-Fock approximation (HF), $n_{\alpha\alpha'}=\delta_{\alpha\alpha'}$ for hole states and 
$n_{\alpha\alpha'}=0$ for particle states, and $C_{\alpha\beta\alpha'\beta'}=0$. 
Keeping in Eq. (\ref{hrpa}) only  the amplitudes $x^\mu_{h}$ and $X^\mu_{hh':p}$, 
where $p$ and $h$ refer to a particle state and a hole state, respectively, 
corresponds to the TDA equation of odd particle systems \cite{RS}. However, even 
within the HF-ground state, Eq.~(\ref{hrpa}) can have all components of 
$X^\mu_{\alpha\beta:\gamma}$; 
$X^\mu_{hh':p}$, $X^\mu_{pp':h}$, $X^\mu_{hh':h''}$, $X^\mu_{pp':p''}$, $X^\mu_{hp:p'}$ 
and $X^\mu_{hp:h'}$.
Such equations have been proposed for the first time in Ref. \cite{hrpa73} and 
they have been applied in Ref. \cite{hrpa96}.
This very much extended configuration space actually leads to some 
difficulties which have been discussed in Ref. \cite{hrpa96}. We will take up this 
discussion again below.

In the following we discuss the relation of $x_\alpha^\mu$ with 
$n_{\alpha\alpha}$. Using Eq. (\ref{stddm1}) for $x^\mu_{\alpha'}$
and the complex conjugate of Eq. (\ref{stddm1}) for $x^\mu_{\alpha}$, we can 
eliminate $\omega_\mu$ obtaining an equation
for $\sum_\mu x^\mu_{\alpha'}(x^\mu_{\alpha})^*$
\begin{eqnarray}
(\epsilon_\alpha&-&\epsilon_{\alpha'})\sum_\mu x^\mu_{\alpha'}(x^\mu_{\alpha})^*
\nonumber \\
&+&\sum_{\lambda_1\lambda_2\lambda_3}[\langle\alpha\lambda_3|v|\lambda_1\lambda_2\rangle
\sum_\mu x^\mu_{\alpha'}(X^\mu_{\lambda_1\lambda_2:\lambda_3})^*
\nonumber \\
&-&\langle\lambda_1\lambda_2|v|\alpha'\lambda_3\rangle
\sum_\mu X^\mu_{\lambda_1\lambda_2:\lambda_3}(x^\mu_{\alpha})^*]
=0.
\label{xx}
\end{eqnarray}
On the other hand the stationary condition $\dot{n}_{\alpha\alpha'}=0$ for Eq. (\ref{tddm1}) gives \cite{Toh}
\begin{eqnarray}
(\epsilon_\alpha&-&\epsilon_{\alpha'})n_{\alpha\alpha'}
\nonumber \\
&+&\sum_{\lambda_1\lambda_2\lambda_3}[\langle\alpha\lambda_3|v|\lambda_1\lambda_2\rangle
C_{\lambda_1\lambda_2\alpha'\lambda_3}
\nonumber \\
&-&\langle\lambda_1\lambda_2|v|\alpha'\lambda_3\rangle
C_{\alpha\lambda_3\lambda_1\lambda_2}]=0.
\label{nn}
\end{eqnarray}
Equations (\ref{xx}) and (\ref{nn}) suggest that $\sum_\mu x^\mu_{\alpha'}(x^\mu_{\alpha})^*$
and $\sum_\mu X^\mu_{\alpha'\beta':\beta}(x^\mu_{\alpha})^*$ correspond to $n_{\alpha\alpha'}$ 
and $C_{\alpha\beta\alpha'\beta'}$, respectively, though the symmetry of $C_{\alpha\beta\alpha'\beta'}$
under the exchange of $\alpha$ and $\beta$
is lost in $\sum_\mu X^\mu_{\alpha'\beta':\beta}(x^\mu_{\alpha})^*$. We will show below that 
$n_{\alpha\alpha}=\sum_\mu x^\mu_\alpha(x^\mu_{\alpha})^*$ approximately holds in the applications to $^{16}$O.

\subsection{Equation of motion approach with excitation operator}
\begin{figure} 
\begin{center} 
\includegraphics[height=6cm]{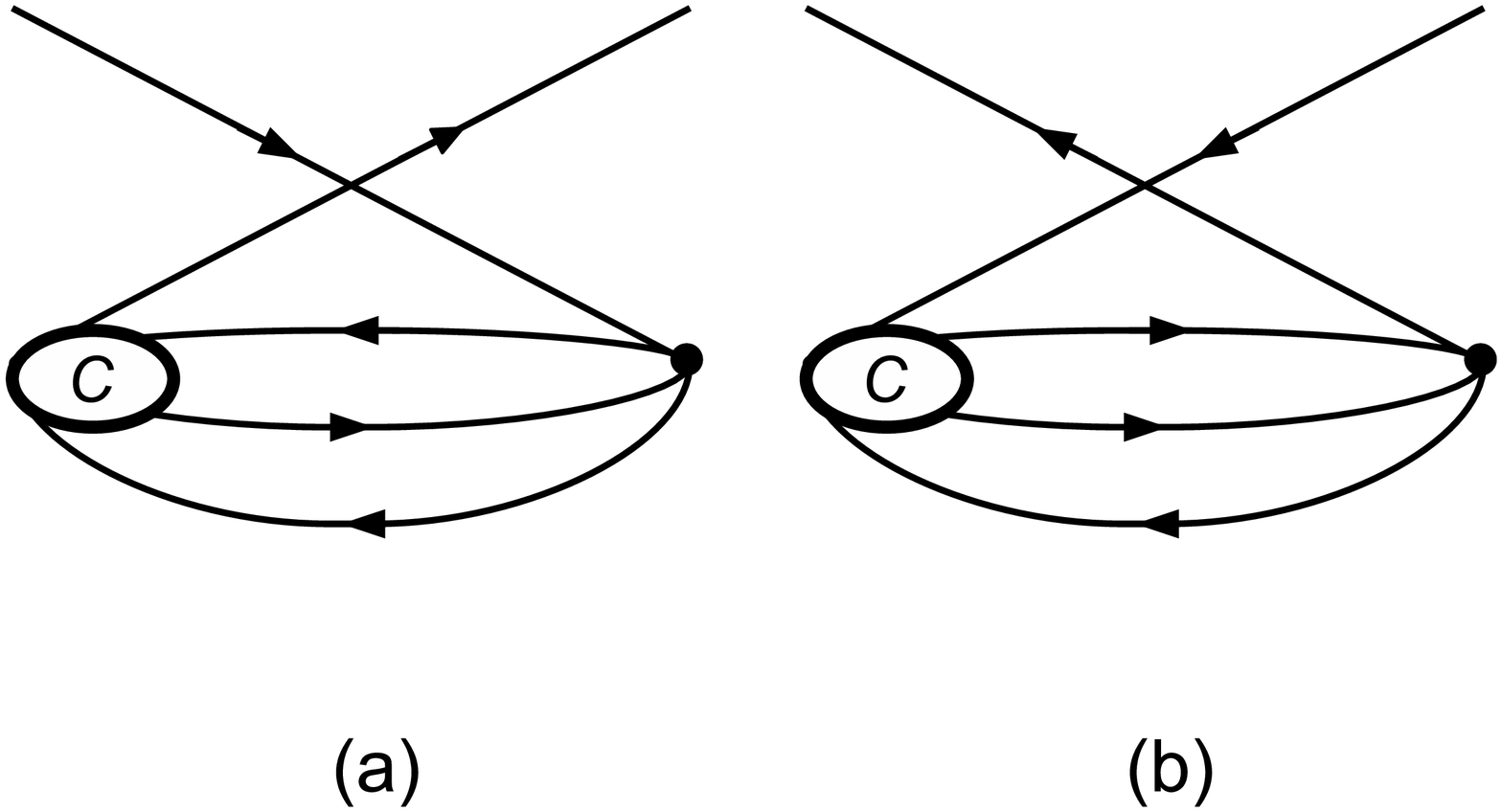}
\end{center}
\caption{(a) Self-energy contribution to a particle state and 
(b) that to a hole state. The ellipses denote $C_{\alpha\beta\alpha'\beta'}$ 
and the dots the residual interaction.}
\label{diagram1} 
\end{figure}

Equation (\ref{hrpa}) lacks some effects such as self-energy contributions in the configuration $X^\mu_{\alpha\beta:\gamma}$,
which should be included when a correlated ground state is used.
In order to take account of such effects, we present another formulation which is based on EoM \cite{Rowe}.
Introducing the excitation operator $q^+_\mu$  
\begin{eqnarray}
q^+_\mu=\sum_\alpha y_\alpha^\mu~ a_\alpha 
+\sum_{\alpha\beta\gamma}Y_{\alpha\beta:\gamma}^\mu:a^+_\gamma a_\beta a_\alpha:
\label{operh}
\end{eqnarray}
and assuming, as usual, $q^+_\mu|0\rangle=|\mu\rangle$ and $q_\mu|0\rangle=0$ 
(for the existence of such a relation, see below),
we obtain from Eqs. (\ref{eq1}) and (\ref{eq2})
\begin{eqnarray}
\left(
\begin{array}{cc}
A&C\\
B&D
\end{array}
\right) 
\left(
\begin{array}{c}
y^{\mu}\\
Y^{\mu}
\end{array}
\right)
=\omega_\mu
\left(
\begin{array}{cc}
N_{11}&N_{12}\\
N_{21}&N_{22}
\end{array}
\right)
\left(
\begin{array}{c}
y^{\mu}\\
Y^{\mu}
\end{array}
\right),
\label{hrpa1}
\end{eqnarray}
where the matrices are defined as
\begin{eqnarray}
A(\alpha:\alpha')&=&\langle 0|\{[H,a^+_{\alpha}],a_{\alpha'}\}|0\rangle, \\
B(\alpha\beta\gamma:\alpha') &=&\langle 0|\{[H,:a^+_\alpha a^+_\beta a_\gamma:], a_{\alpha'}\}|0\rangle,\\
C(\alpha:\alpha'\beta'\gamma')&=&\langle 0|\{[H,a^+_{\alpha}:],:a^+_{\gamma'} a_{\beta'}a_{\alpha'} \}|0\rangle,
\nonumber \\
D(\alpha\beta\gamma:\alpha'\beta'\gamma'&)&
\nonumber \\
=\langle 0|\{[H&,&:a^+_\alpha a^+_\beta a_\gamma:],:a^+_{\gamma'} a_{\beta'} a_{\alpha'}:\}|0\rangle,
\end{eqnarray}
\begin{eqnarray}
N_{11}(\alpha:\alpha')&=&\langle 0|\{a^+_\alpha, a_{\alpha'}\}|0\rangle=\delta_{\alpha\alpha'}, \\
N_{12}(\alpha:\alpha'\beta'\gamma)&=&\langle 0|\{a^+_\alpha, :a^+_{\gamma'} a_{\beta'} a_{\alpha'}:\}|0\rangle=0,\\
N_{21}(\alpha\beta\gamma:\alpha')&=&\langle 0|\{:a^+_\alpha a^+_\beta a_\gamma:, a_{\alpha'}\}|0\rangle=0,
\nonumber \\
N_{22}(\alpha\beta\gamma:\alpha'\beta'\gamma'&)&
\nonumber \\
=\langle 0|\{:a^+_\alpha a^+_\beta a_\gamma&,&:a^+_{\gamma'} a_{\beta'} a_{\alpha'}:\}|0\rangle.
\label{norm22}
\end{eqnarray}
Here $\{~\}$ implies the anticommutator, $\{A,B\}=AB+BA$.
The norm matrix $N_{22}$ is given in Appendix A.
The matrix elements in Eq.~(\ref{hrpa1})
can be expressed using those in Eq.~(\ref{hrpa}) such as
\begin{eqnarray}
A=a\times N_{11}\\
B=b\times N_{11},\\
C=c\times N_{22}.
\end{eqnarray}
The matrix $D$ consists of the two types of terms, one expressed by 
$D_1=d\times N_{22}$ and the other given by $D_2$, which originates from the terms with $:a^+_{\lambda_5} a^+_{\lambda_4} a^+_{\lambda_3} a_{\lambda_2} a_{\lambda_1}:$ 
in $[H,:a^+_\alpha a^+_\beta a_\gamma:]$:
\begin{eqnarray}
 &[&H,:a^+_\alpha a^+_\beta a_\gamma:]=\sum_{\alpha'}c(\alpha\beta\gamma:\lambda)a^+_{\alpha'}
\nonumber \\
 &+&\sum_{\alpha'\beta'\gamma'}d(\alpha\beta\gamma:\alpha'\beta'\gamma'):a^+_{\alpha'}a^+_{\beta'}a_{\gamma'}:
\nonumber \\ 
 &+&\sum_{\lambda_1\lambda_2\lambda_3\lambda_4\lambda_5}e(\alpha\beta\gamma:\lambda_1\lambda_2\lambda_3\lambda_4\lambda_5)
 \nonumber \\
&\times&
:a^+_{\lambda_1}a^+_{\lambda_2}a^+_{\lambda_3}a_{\lambda_5}a_{\lambda_4}:.
\label{d2}
\end{eqnarray}
Using 
\begin{eqnarray}
N_{32}&(&\lambda_1\lambda_2\lambda_3\lambda_4\lambda_5:\alpha\beta\gamma)
\nonumber \\
&=&
\langle 0|\{:a^+_{\lambda_1} a^+_{\lambda_2} a^+_{\lambda_3} a_{\lambda_5} a_{\lambda_4}:, :a^+_\gamma a_\beta a_\alpha:\}|0\rangle,
\end{eqnarray}
$D_2$ can be expressed as $e\times N_{32}$.
These terms include, for example,
\begin{eqnarray}
-\frac{1}{2}\delta_{\alpha\alpha'}\delta_{\beta\beta'}\sum_{\lambda_1\lambda_2\lambda_3}
\langle\lambda_1\lambda_2|v|\gamma'\lambda_3\rangle_A C_{\gamma\lambda_3\lambda_1\lambda_2},
\nonumber
\end{eqnarray}
which is a self-energy contribution to the state $\gamma$. The self-energy contributions are schematically shown in
Fig. \ref{diagram1}.

The normalization of the amplitudes is given by 
\begin{eqnarray}
\left(y^{\mu~*}~Y^{\mu~*}\right)
\left(
\begin{array}{cc}
N_{11}&N_{12}\\
N_{21}&N_{22}
\end{array}
\right)
\left(
\begin{array}{c}
y^{\mu'}\\
Y^{\mu'}
\end{array}
\right)=\delta_{\mu\mu'}.
\label{normalization}
\end{eqnarray}
The closure relation is written as 
\begin{eqnarray}
\sum_\mu\left(
\begin{array}{c}
y^{\mu}\\
Y^{\mu}
\end{array}
\right)\left(y^{\mu~*}~Y^{\mu~*}\right)
\left(
\begin{array}{cc}
N_{11}&N_{12}\\
N_{21}&N_{22}
\end{array}
\right)
=I,
\label{closure}
\end{eqnarray}
where $I$ is the unit matrix.
We refer to the formulation Eq.~(\ref{hrpa1}) as the extended odd-RPA (EoRPA).

In the following we discuss the relation between Eqs. (\ref{hrpa}) 
and (\ref{hrpa1}).
The transition amplitudes $x^\mu_{\alpha}$ and $X^\mu_{\alpha\beta:\gamma}$ 
are given by $y^\mu_{\alpha}$ and $Y^\mu_{\alpha\beta:\gamma}$ as 
\begin{eqnarray} 
\left(
\begin{array}{c}
x^{\mu}\\
X^{\mu}
\end{array}
\right)
=
\left(
\begin{array}{cc}
N_{11}&N_{12}\\
N_{21}&N_{22}
\end{array}
\right)
\left(
\begin{array}{c}
y^{\mu}\\
Y^{\mu}
\end{array}
\right).
\end{eqnarray}
Inserting this expression into Eq. (\ref{hrpa}), we obtain
\begin{eqnarray}
\left(
\begin{array}{cc}
A&C\\
B&D_1
\end{array}
\right) 
\left(
\begin{array}{c}
y^{\mu}\\
Y^{\mu}
\end{array}
\right)
=\omega_\mu
\left(
\begin{array}{cc}
N_{11}&N_{12}\\
N_{21}&N_{22}
\end{array}
\right)
\left(
\begin{array}{c}
y^{\mu}\\
Y^{\mu}
\end{array}
\right).
\label{hrpa2}
\end{eqnarray}
The difference between Eqs. (\ref{hrpa1}) and (\ref{hrpa2}) and thus between Eqs. (\ref{hrpa1}) and (\ref{hrpa}) 
resides in the matrix $D$. Some effects of the ground-state correlations
such as the self-energy contributions are missing in Eq. (\ref{hrpa2}) and thus in Eq. (\ref{hrpa}) as mentioned above. The importance of these missing terms will be discussed below in the application section.

\subsubsection{Symmetry properties}
First we show that the Hamiltonian matrix of Eq.~(\ref{hrpa1}) is hermitian.
We use the operator identity
\begin{eqnarray}
\langle 0|\{[H,\hat{A}],\hat{B}\}|0\rangle&+&\langle 0|\{[H,\hat{B}],\hat{A}\}|0\rangle
\nonumber \\
&=&\langle 0|[H,\{\hat{A},\hat{B}\}]|0\rangle.
\label{jacobi}
\end{eqnarray}
In Eq.~(\ref{hrpa1}), in the matrix $A$, the operators $\hat{A}$ and $\hat B$ 
are identified with $a^+_\alpha$ and $a_{\alpha'}$, respectively. Since
$\{\hat{A},\hat{B}\}$ is unity, the right-hand side
of Eq.~(\ref{jacobi}) vanishes, which means $\langle 0|\{[H,\hat{A}],\hat{B}\}|0\rangle=-\langle 0|\{[H,\hat{B}],\hat{A}\}|0\rangle$
and 
\begin{eqnarray}
A(\alpha:\alpha')^* &=&-\langle 0|\{[H,a_\alpha], a^+_{\alpha'}\}|0\rangle,
\nonumber \\
&=&\langle 0|\{[H,a^+_{\alpha'}],a_\alpha\}|0\rangle
=A(\alpha':\alpha).
\end{eqnarray}
In the case of the matrix $B$ in Eq.~(\ref{hrpa1}) $\hat{A}$ is $:a^+_\alpha a^+_\beta a_\gamma:$ and $\hat{B}$ is $a_{\alpha'}$, and
$\{\hat{A},\hat{B}\}$ is reduced to a one-body operator. Due to the ground-state condition Eq.~(\ref{tddm1}) the right-hand side
of Eq.~(\ref{jacobi}) vanishes, which means $\langle 0|\{[H,\hat{A}],\hat{B}\}|0\rangle=-\langle 0|\{[H,\hat{B}],\hat{A}\}|0\rangle$
and 
\begin{eqnarray}
B(\alpha\beta\gamma:\alpha')^* &=&-\langle 0|\{[H,:a^+_\gamma a_\beta a_\alpha :], a^+_{\alpha'}\}|0\rangle,
\nonumber \\
&=&\langle 0|\{[H,a^+_{\alpha'}], :a^+_\gamma a_\beta a_\alpha:\}|0\rangle
\nonumber \\
&=&C(\alpha':\alpha\beta\gamma).
\end{eqnarray}
Similarly, for the matrix $D$ in Eq.~(\ref{hrpa1}) $\hat{A}$ is $:a^+_\alpha a^+_\beta a_\gamma:$ and $\hat{B}$ is
$:a^+_{\gamma'} a_{\beta'} a_{\alpha'}$, and $\{\hat{A},\hat{B}\}$ is reduced to at most a two-body operator. Due to the
ground-state conditions Eqs. (\ref{tddm1}) and (\ref{tddm2}) the right-hand side of Eq.~(\ref{jacobi}) vanishes, which implies
\begin{eqnarray}
D(\alpha\beta\gamma:\alpha'\beta'\gamma')^*=D(\alpha'\beta'\gamma':\alpha\beta\gamma).
\end{eqnarray}
Therefore, the Hamiltonian matrix in Eq. (\ref{hrpa1}) is hermitian.
In the applications, shown below, we do not take all the matrix elements of $n_{\alpha\alpha'}$ and $C_{\alpha\beta\alpha'\beta'}$,
which causes a violation of the hermiticity of Eq.~(\ref{hrpa1}), though 
it will turn out to be small.

Next we discuss the relation between the formulations for a hole state and a particle state.
We can obtain a formulation for a particle state using the excitation operator
\begin{eqnarray}
q^+_\mu=\sum_\alpha z_\alpha^\mu ~a^+_\alpha 
+\sum_{\alpha\beta\gamma}Z_{\alpha\beta:\gamma}^\mu:a^+_\alpha a^+_\beta a_\gamma:.
\end{eqnarray}
Since this operator is the conjugate of the hole-state excitation operator Eq.~(\ref{operh}), it is easily shown that
the formulation for a particle state is given as
\begin{eqnarray}
\left(
\begin{array}{cc}
A&C\\
B&D
\end{array}
\right)^t
\left(
\begin{array}{c}
z^{\mu}\\
Z^{\mu}
\end{array}
\right)
=\omega_\mu
\left(
\begin{array}{cc}
N_{11}&N_{12}\\
N_{21}&N_{22}
\end{array}
\right)^t
\left(
\begin{array}{c}
z^{\mu}\\
Z^{\mu}
\end{array}
\right),
\label{hrpap}
\end{eqnarray}
where the superscript $t$ means the transposition of the corresponding matrix and $\omega_\mu$ is defined
by $\omega_\mu=E_\mu-E_0$.
Equation (\ref{hrpap}) implies that $(z^\mu,~Z^\mu)$ is the left-hand eigenvector of Eq.~(\ref{hrpa1}).
Thus Eq.~(\ref{hrpa1}) gives simultaneously the particle states and the hole states. 
This is completely analogous to pp(hh)RPA (see Ref. \cite{RS}.)

\subsubsection{Hartree-Fock approximation for the ground state}
\begin{figure} 
\begin{center} 
\includegraphics[height=6cm]{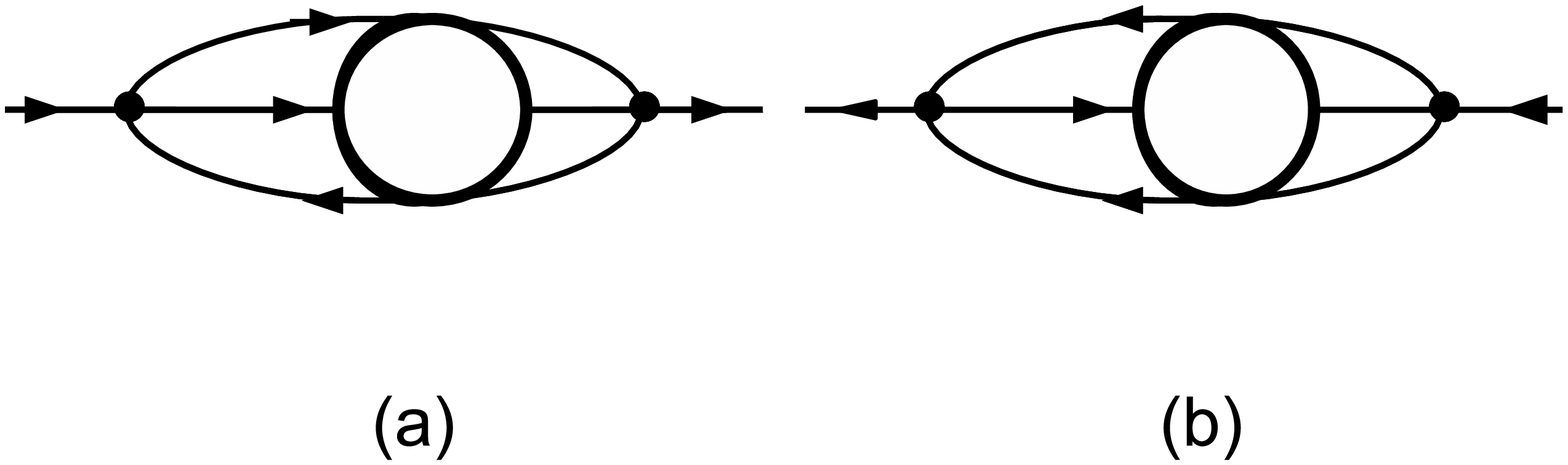}
\end{center}
\caption{(a) Mass operator for a particle state described by $Y^\mu_{pp':h}$ and
(b) that for a hole state described by $Y^\mu_{hh':p}$. The circles mean the propagators given by $Y^\mu_{pp':h}$ ((a))
and $Y^\mu_{hh':p}$ ((b)), and the dots the residual interaction. }
\label{diagram2} 
\end{figure}
\begin{figure} 
\begin{center} 
\includegraphics[height=6cm]{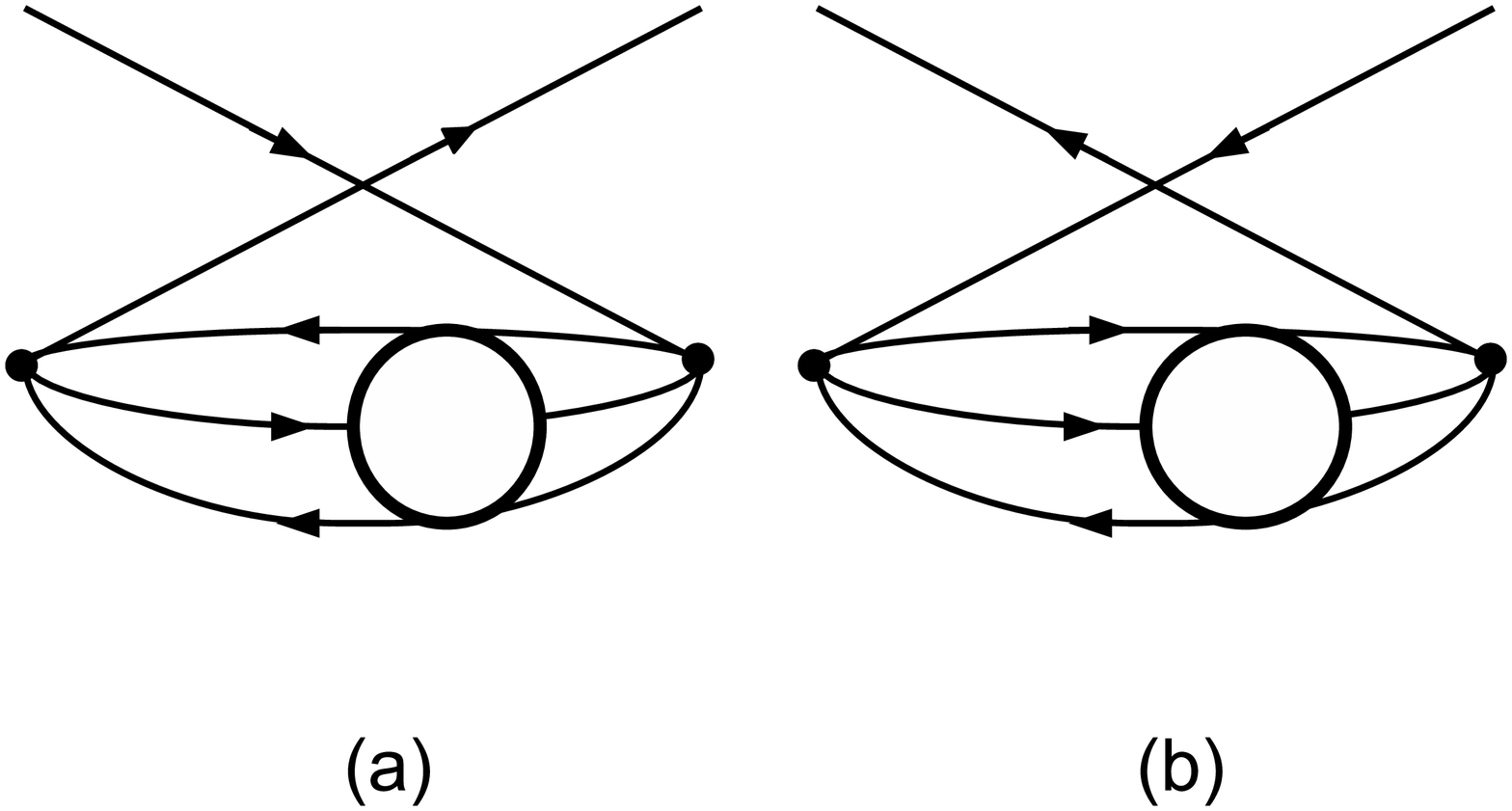}
\end{center}
\caption{(a) Mass operator for a particle state described by $Y^\mu_{hh':p}$ and 
(b) that for a hole state described by $Y^\mu_{pp':h}$. The circles mean the propagators given by $Y^\mu_{hh':p}$ ((a))
and $Y^\mu_{pp':h}$ ((b)), and the dots the residual interaction.}
\label{diagram3} 
\end{figure}

If we make the usual approximation to take for the ground state $|0\rangle$ 
the HF one, 
$N_{22}$ in Eq. (\ref{norm22}) becomes
\begin{eqnarray}
N_{22}(\alpha\beta\gamma&:&\alpha'\beta'\gamma')=
(\delta_{\alpha\alpha'}\delta_{\beta\beta'}-\delta_{\alpha\beta'}\delta_{\beta\alpha'})\delta_{\gamma'\gamma}
\nonumber \\
&\times&(n^0_{\gamma\gamma}+n^0_{\alpha\alpha}n^0_{\beta\beta}-n^0_{\gamma\gamma}n^0_{\alpha\alpha}-n^0_{\gamma\gamma}n^0_{\beta\beta}),
\end{eqnarray}
where $n^0_{\alpha\alpha}$ is equal to 1 or 0. In HF, $N_{22}$ is non-vanishing only for $Y_{pp':h}^\mu$ and $Y_{hh':p}^\mu$. 
These amplitudes $Y_{pp:h}^\mu$ and $Y_{hh':p}^\mu$ correspond to the backward amplitudes of $y_h^\mu$ and $y_p^\mu$, respectively.
Hereafter we refer to this formulation consisting of the four amplitudes, 
$y_h^\mu$, $y_p^\mu$, $Y_{pp:h}^\mu$ and 
$Y_{hh':p}^\mu$  
as odd-RPA (oRPA). 
The mass operators of the one-body Green's function derived from oRPA
are schematically shown in Figs. \ref{diagram2} and \ref{diagram3}. 
Since oRPA describes the hole states and particle states simultaneously, 
the single-particle strength can be spread over both positive and negative energy regions.
We consider that the strength below the Fermi energy $\epsilon_{F}$ of the core nucleus belongs to the states in the $A-1$ system,
while that above $\epsilon_{F}$ to the states in the $A+1$ system.
The TDA hole-state equation is obtained by keeping only $y^\mu_h$ and $Y^\mu_{hh':p}$:
since the coupling of $y^\mu_h$ to $Y^\mu_{hh':p}$ is included in addition to the coupling to the backward amplitude $Y_{pp':h}^\mu$,
our oRPA actually corresponds to some sort of second RPA for even nucleon systems.
One may think that in addition the amplitudes $Y^\mu_{hp:h'}$ and $Y^\mu_{pp':p''}$ should be included in oRPA because they respectively express
the backward propagations of the particle - hole pair and the hole-hole pair in $Y^\mu_{hh':p}$. However, these amplitudes
cannot be included because the norm of these amplitudes is not defined in HF
(the matrix elements $N_{22}$ for $Y^\mu_{hp:h'}$ and $Y^\mu_{pp':p''}$ vanish in HF).
As mentioned above, the formulation Eq. (\ref{hrpa}) allows us to implement all the $X^\mu_{\alpha\beta:\gamma}$ amplitudes
including $X^\mu_{hp:h'}$ and $X^\mu_{pp':p''}$ because there is no restriction of the norm matrix.
(If Eq. (\ref{hrpa2}) is used instead of Eq. (\ref{hrpa}), $X^\mu_{hp:h'}$ and $X^\mu_{pp':p''}$ are projected out, however.) 
The inclusion of all the amplitudes of $X^\mu_{\alpha\beta:\gamma}$ can give quite
unphysical results because the sum of some unperturbed energies 
corresponding to $X^\mu_{\alpha\beta:\gamma}$ fall near $\epsilon_F$, which
makes it difficult to distinguish the hole states from the particle states
(see also Ref. \cite{hrpa96}).
For these reasons we mainly present the results in EoRPA calculated using 
only the four amplitudes corresponding to $y_h^\mu$, $y_p^\mu$, $Y_{pp':h}^\mu$ and 
$Y_{hh':p}^\mu$ in oRPA, although the matrix 
elements of $N_{22}$ are nonvanishing for all configurations due to the ground 
state correlations and, therefore, all other $Y^\mu_{\alpha\beta:\gamma}$ could 
be included, in principle. We investigate the effect of inclusion of those other amplitudes in EoRPA 
in some limited cases.

\subsubsection{The RPA ground state wave function}
The choice of the subspace spanned by the afore mentioned four amplitudes 
$y_h^{\mu}$, $y_p^{\mu}$, $Y_{pp':h}^{\mu}$, $Y_{hh':p}^\mu$ may be given a different rationale.
We consider the following two quasi-particle operators which consist only of the forward and backward amplitudes 
\begin{eqnarray}
q^+_{\alpha} &=&\sum_py_p^{\alpha} a^+_p -\frac{1}{2} \sum_{hh'p}Y^{\alpha}_{hh':p}
a^+_ha^+_{h'}a_p,
\label{aurpa1}
\\
q^+_{\rho} &=& \sum_h y_h^{\rho} a_h - \frac{1}{2} \sum_{pp'h} Y^{\rho}_{pp':h}
a_h^+a_pa_{p'}
\label{aurpa2}
\end{eqnarray}
and neglect the coupling of $y^{\alpha}_p$ to $Y^\alpha_{pp':h}$ and that of $y_h^{\rho}$ to $Y^{\rho}_{hh':p}$.
This oRPA scheme actually corresponds to the standard RPA for even nucleon systems. This can for example be seen in the following way.
It can easily be shown that the operators $q_{\alpha}$ and $q_{\rho}$ 
kill the following RPA vacuum, i.e. $q|Z\rangle=0$ with
\begin{eqnarray}
|Z\rangle = e^{\frac{1}{4}\sum z_{pp'hh'}a^+_p a_h a^+_{p'} a_{h'}}|{\rm HF}\rangle
\label{gsfunc}
\end{eqnarray}
under the conditions
\begin{eqnarray}
\sum_p y^{\alpha *}_p z_{pp'hh'} &=& Y^{\alpha *}_{hh':p'}
\\
\sum_h y^{\rho *}_h z_{pp'hh'} &=& Y^{\rho *}_{pp':h'},
\end{eqnarray}
where $|{\rm HF}\rangle$ is the HF ground state of an even $A$ system.

These two quasiparticles (one for the particle addition ($\alpha$) and one for the 
particle removal ($\rho$)) span, as seen, exactly the space of the four amplitudes 
discussed in II.B.2. However, the single equation for the four amplitudes is 
now split into two independent $2\times 2$ equations corresponding to the two 
operators introduced 
in Eqs. (\ref{aurpa1}) and (\ref{aurpa2}), respectively. Using in these equations the HF ground 
state as in II.B.2, we see, that we have one type of 'forward' going 
amplitudes and one type of 
'backward going' amplitudes in analogy with what we know from 
standard ph RPA for 
even systems with corresponding amplitudes $X$ and $Y$. 
As a matter of fact, it recently has been shown \cite{jemai}
that also for the standard ph-RPA a generalized operator can be found which annihilates the state Eq. (\ref{gsfunc}). 
It is given by the following form
\begin{eqnarray}
Q_{\nu}&=&\sum_{ph}[X_{ph}^{\nu*}a_h^+a_p- Y_{ph}^{\nu*}a_p^+a_h]
\nonumber \\
&+&\frac{1}{2}\sum_{php_1p_2}\eta^{\nu}_{p_1p_2ph}a_{p_2}^+a_{p_1}a_p^+a_h
\nonumber \\
&-&\frac{1}{2}\sum_{phh_1h_2}\eta^{\nu}_{h_1h_2ph}a_{h_1}^+a_{h_2}a_p^+a_h.
\end{eqnarray}
This destruction operator kills the vacuum Eq. (\ref{gsfunc}), i.e. $Q|Z\rangle=0$, under the conditions
\begin{eqnarray}
z_{php'h'}&=&\sum_{\nu}(X^{-1})^{\nu}_{ph}Y^{\nu}_{p'h'}
\\
\eta^{\nu}_{p_1p_2ph}&=&\frac{1}{2}\sum_{h_1}X^{\nu}_{p_1h_1}z_{pp_2hh_1}
\\
\eta^{\nu}_{h_1h_2ph}&=&\frac{1}{2}\sum_{p_1}X^{\nu}_{p_1h_1}z_{pp_1hh_2}.
\end{eqnarray}
We see that there are additional terms to the standard ph-RPA operator which 
contain specific two-body terms. The corresponding terms in $Q_\nu^+$ can 
schematically be obtained in augmenting 
the addition operator Eq. (\ref{aurpa1}) by a destructor $a_h$ and the removal operator Eq. (\ref{aurpa2}) by 
a creator $a^+_p$. The $\eta$-terms are also small amplitude 
(backward going) terms which can be added to the standard RPA, evaluated  
with the HF state. They improve the results of standard RPA \cite{jemai1}. We, 
therefore, see that complete consistency between RPA in even and odd 
systems can be achieved.

\subsubsection{Green's Function Description}

It may be instructive to cast the above amplitude equations into Green's function language. For this we write down a Dyson equation
\begin{eqnarray}
G^{\omega}_{kk'} = G^0_k\delta_{kk'} + G^0_k\sum_{k_1}M^{\omega}_{kk_1}G^{\omega}_{k_1k'},
\end{eqnarray}
where
\begin{eqnarray}
G^0_k = \frac{1 - n^{(0)}_k}{\omega - \varepsilon_k + i \eta}+
\frac{n^{(0)}_k}{\omega - \varepsilon_k - i \eta}
\end{eqnarray}
is the free or HF Green's function with the occupation numbers $n^{(0)}_k$ equal to 0 or 1.
The mass operator is given by
\begin{eqnarray}
M_{kk'} &=& \sum_{\alpha hh'p_1p_2p'_1p'_2} \langle kh|v|p_1p_2\rangle
\frac{Y^{\rho}_{p_1p_2:h}Y^{\rho *}_{p_1'p_2':h'}}{\omega - \Omega_{\rho}^{N+1} + i\eta}
\nonumber \\
&\times&\langle p_1'p_2'|v|h'k'\rangle
\nonumber \\
&+&
\sum_{\rho pp'h_1h_2h'_1h'_2}\langle kp|v|h_1h_2\rangle
\frac{Y^{\alpha}_{h_1h_2:p}Y^{\alpha *}_{h_1'h_2':p'}}{\omega - \Omega_{\alpha}^{N-1} - i\eta}
\nonumber \\
&\times&\langle h_1'h_2'|v|p'k'\rangle,
\end{eqnarray}
where $Y^{\alpha,\rho}$ and $\Omega_{\alpha,\rho}$ are the TDA 2p-1h and 2h-1p amplitudes and eigenvalues, respectively, 
obtained from the corresponding TDA equations \cite{RS}. In the case where we strictly work with oRPA corresponding 
to the ground state Eq. (\ref{gsfunc}), the coupled system of particle and hole propagation in above Dyson equation 
decouples into two separate Dyson equations, one for the particles (with the $\alpha$ part of the mass operator corresponding to Fig. \ref{diagram3}(a)) 
and one for the holes (with the $\rho$ part of the mass operator corresponding to Fig. \ref{diagram3}(b)). 
It may certainly be appealing to work with an approach which is based on a ground state wave function.\\

Going beyond the use of a HF ground state, we can do as in this work considering EoRPA as described above. 
However, there is also the possibility to mix even and odd RPA's. For example it has turned out that Self Consistent RPA (SCRPA) based on the vacuum Eq. (\ref{gsfunc}) 
gives very good results \cite{jemai}. For instance, it also solves the two particle case exactly. One thus could use SCRPA 
to calculate the correlation functions appearing in an extended oRPA. To use the ans\"atze 
Eqs. (\ref{aurpa1}) and (\ref{aurpa2}) directly seems difficult, since they correspond to a nonlinear 
transformation among the fermion operators.

\subsubsection{Spurious modes}

First we discuss an RPA-like formulation that can bring the c.o.m motion of an odd system at zero excitation energy. 
We consider for an $A+1$ system the ground state $|\Phi_0\rangle$ and an excited state $|\Phi_\mu \rangle$ with excitation energy
$\omega_\mu$.
Using the equation of motion
\begin{eqnarray}
\langle\Phi_0|[a^+_{\alpha'} a_{\alpha},H]|\Phi_\mu\rangle&=&\omega_\mu\langle\Phi_0|a^+_{\alpha'} a_{\alpha}|\Phi_\mu\rangle
\nonumber \\
&=&\omega_\mu x^\mu_{\alpha\alpha'}
\end{eqnarray}
and assuming 
$|\Phi_0\rangle=a^+_p|{\rm HF}\rangle$,
we obtain the following equation 
\begin{eqnarray}
\omega_\mu x^\mu_{\alpha\alpha'}&=&(\epsilon_{\alpha}-\epsilon_{\alpha'})x^\mu_{\alpha\alpha'}
\nonumber \\
&-&(n^0_{\alpha\alpha}-n^0_{\alpha'\alpha'})\sum_{\lambda\lambda'}\langle\alpha\lambda'|v|\alpha'\lambda\rangle_Ax^\mu_{\lambda\lambda'}
\nonumber \\
&+&\sum_{\lambda\lambda'}[\delta_{\alpha p}\langle p \lambda'|v|\alpha'\lambda\rangle_A 
-\delta_{\alpha' p}\langle \alpha \lambda'|v|p\lambda\rangle_A]x^\mu_{\lambda\lambda'}
\nonumber \\
&+&\sum_\lambda [\langle \alpha p|v|\lambda p\rangle_Ax^\mu_{\lambda\alpha'}
-\langle \lambda p|v|\alpha' p\rangle_Ax^\mu_{\alpha\lambda}].
\label{rpasp}
\end{eqnarray}
The first two lines of the above equation have the same form as the standard RPA for an even $A$ system, and the third and fourth
terms are due to the additional nucleon in a particle state $p$.
For the total momentum operator
\begin{eqnarray} 
{\bm P}=\sum_{\alpha\alpha'}\langle\alpha'|-i\hbar\nabla|\alpha\rangle a^+_{\alpha'} a_{\alpha},
\end{eqnarray}
which satisfies $[{\bm P},H]=0$,
we evaluate $\omega_\mu \langle\Phi_0|{\bm P}|\Phi_\mu\rangle$ as
\begin{eqnarray}
\omega_\mu \langle\Phi_0| {\bm P}|\Phi_\mu\rangle=\sum_{\alpha\alpha'}\langle\alpha'|-i\hbar\nabla|\alpha\rangle
\omega_\mu x^\mu_{\alpha\alpha'}.
\end{eqnarray}
Using the right-hand side of Eq. (\ref{rpasp}) and the translational invariance of the interaction \cite{ts04}, we can show
$\omega_\mu \langle\Phi_0|{\bm P}|\Phi_\mu\rangle=0$, which implies $\omega_\mu=0$.
Thus the excitation energy of the c.o.m motion of an odd system given by Eq. (\ref{rpasp}) is zero from the ground state $|\Phi_0\rangle$ and $\epsilon_{p}$ from 
$|\rm{HF}\rangle$. In order to obtain this conclusion, however,
we need to include all components of $x^\mu_{\alpha\alpha'}$ because of the last two terms on the right-hand side of Eq. (\ref{rpasp}).

Now we discuss the c.o.m of a core nucleus in odd $A$ nuclei whose treatment is 
of particular 
relevance. In the standard particle vibration coupling model \cite{bortignon,Giai} the 
spurious mode is simply discarded, first, for the translational mode, on 
physical grounds but also because the RPA amplitudes of a zero mode cannot 
be normalized. On the other hand, e.g. in the case of rotations, it would be 
very important to find a way to include the rotational mode, since it is a 
physical state.
In order to learn something about the 
coupling of single-particle motion and recoil of the core nucleus,
we first show that $\omega_\mu\langle\mu|{\bm P}a_\alpha|0\rangle=\epsilon_\alpha\langle\mu|a_\alpha|0\rangle$ holds in the mean-field approximation.
Using the complex conjugate of Eq. (\ref{eq2}), we evaluate $\omega_\mu\langle\mu|{\bm P}a_\alpha|0\rangle$ such that
\begin{eqnarray}
\omega_\mu\langle\mu|{\bm P}a_\alpha|0\rangle
&=&\langle\mu|[{\bm P}a_\alpha,H]|0\rangle
\nonumber \\
&=&\langle\mu|[{\bm P},H]a_\alpha|0\rangle+\langle\mu|{\bm P}[a_\alpha,H]|0\rangle
\nonumber \\
&=&\langle\mu|{\bm P}[a_\alpha,H]|0\rangle,
\end{eqnarray}
where we use $[{\bm P},H]=0$. If we use the mean-field approximation for $[a_\alpha,H]$, that is,
$[a_\alpha,H]=\epsilon_\alpha a_\alpha$,
then we obtain $\omega_\mu\langle\mu|{\bm P}a_\alpha|0\rangle=\epsilon_\alpha\langle\mu|a_\alpha|0\rangle$, which means that
the strength $|\langle \mu|{\bm P}a_\alpha|0\rangle| ^2$ is concentrated at the state with $\omega_\mu=\epsilon_\alpha$.
In the general case the mean-field approximation is not valid as Eq. (\ref{stddm1}) indicates.
In the realistic applications of our oRPA or 
EoRPA approaches shown below, we will, therefore, see that
a large portion of the strength is distributed to an energy region lower 
than $\epsilon_\alpha$,
which can be interpreted as a recoil effect of the core nucleus. \\

In the past, the question of the spurious modes appeared essentially in the particle-vibration coupling model \cite{Giai} 
which is derived from the Green's function method factorizing in the mass operator the 2p-1h (2h-1p) propagator into an 
ph-RPA propagator and a HF single particle propagator. In the spectral representation of the RPA propagator the spurious mode 
is then discarded because of the zero energy mode and the ensuing diverging amplitudes. On the other hand, if one could solve 
the 2p-1h (2h-1p) propagator in the mass operator exactly (e.g. in a model) or with a consistent higher order theory, surely 
no problem with a spurious motion of the core nucleus would be present. 
From our analysis above, it appears that the mass operator should be calculated with 2h-1p (2p-1h) TDA amplitudes. It could very well be that this approach gives more realistic 
results than the particle vibration coupling model where the spurious mode is discarded. That is what our derivation seems to indicate.

In any case, e.g. in the case of rotations, it would be necessary to include this mode, since it is physical. One could push the argument 
even further and assume that, since, e.g. the rotation is very collective, the factorization of the 2h-1p (2p-1h) TDA into 
a ph-TDA + plus a hole (particle) is a good approximation (the neglected terms coming only from exchange). 
Because of its strong collectivity, eventually all the other couplings to intrinsic ph modes could be neglected. 
Actually analogous questions would arise in cold fermionic atom systems where one could ask the question 
what happens to an odd fermion which is coupled to the so-called Kohn mode, i.e. a coherent c.o.m. motion of the underlying even system,
in the external harmonic container. Since the mass of the core can be very large, e.g. with a million of atoms, 
the factorization can become quite valid and also the ph-TDA for the Kohn mode will become very collective. 
It could be interesting to investigate this question 
in more detail theoretically and experimentally because the treatment of Goldstone modes in single-particle 
mass operators is, to the best of our knowledge, an unsolved problem.

\subsection{The $A=2$ case}
We show that our formulation is exact for an $A=2$ system.
In the case of an $A=2$ system, TDDM gives
the coupled equations of motion for $n_{\alpha\alpha'}$
and the two-body density matrix $\rho_{\alpha\beta\alpha'\beta'}$, which are 
defined as
\begin{eqnarray}
n_{\alpha\alpha'}(t)&=&\langle\Phi(t)|a^+_{\alpha'} a_\alpha|\Phi(t)\rangle,
\\
\rho_{\alpha\beta\alpha'\beta'}(t)&=&\langle\Phi(t)|a^+_{\alpha'}a^+_{\beta'}
 a_{\beta}a_{\alpha}|\Phi(t)\rangle,
 \label{rho2}
\end{eqnarray}
where $|\Phi(t)\rangle$ is the time-dependent total wavefunction
$|\Phi(t)\rangle=\exp[-iHt] |\Phi(t=0)\rangle$.
The equations in TDDM are written as \cite{qdot}
\begin{eqnarray}
i \hbar\dot{n}_{\alpha\alpha'}&=&\sum_{\lambda}(\langle\alpha|t|\lambda\rangle{n}_{\lambda\alpha'}
-\langle\lambda|t|\alpha'\rangle{n}_{\alpha\lambda})
\nonumber \\
&+&\sum_{\lambda_1\lambda_2\lambda_3}
[\langle\alpha\lambda_1|v|\lambda_2\lambda_3\rangle \rho_{\lambda_2\lambda_3\alpha'\lambda_1}
\nonumber \\
&-&\rho_{\alpha\lambda_1\lambda_2\lambda_3}\langle\lambda_2\lambda_3|v|\alpha'\lambda_1\rangle],
\label{n2}
\end{eqnarray}
\begin{eqnarray}
i\hbar \dot{\rho}_{\alpha\beta\alpha'\beta'}&=&
\sum_\lambda(\langle\alpha|t|\lambda\rangle\rho_{\lambda\beta\alpha'\beta'}
+\langle\beta|t|\lambda\rangle\rho_{\alpha\lambda\alpha'\beta'}
\nonumber \\
&-&\langle\lambda|t|\alpha'\rangle\rho_{\alpha\beta\lambda\beta'}
-\langle\lambda|t|\beta'\rangle\rho_{\alpha\beta\alpha'\lambda})
\nonumber \\
&+&\sum_{\lambda_1\lambda_2}[
\langle\alpha\beta|v|\lambda_1\lambda_2\rangle\rho_{\lambda_1\lambda_2\alpha'\beta'}
\nonumber \\
&-&\langle\lambda_1\lambda_2|v|\alpha'\beta'\rangle\rho_{\alpha\beta\lambda_1\lambda_2}].
\label{N2C2}
\end{eqnarray}
Here the single-particle states are arbitrary.
Since there are no higher-level reduced density matrices in an $A=2$ system, these two equations are exact.
When the two-body density matrix in Eq. (\ref{n2}) is approximated by anti-symmetrized products of the occupation matrices,
Eq. (\ref{n2}) is equivalent to the equation in the time-dependent HF theory.
The ground state is given as a stationary solution of these equations. 

The equation for the transition amplitude $x^\mu_\alpha$ is 
\begin{eqnarray}
\sum_{\lambda}(\langle\lambda|t|\alpha\rangle -\delta_{\alpha\lambda}\omega_\mu)x^\mu_\lambda
&+&\sum_{\lambda_1\lambda_2\lambda_3}\langle\lambda_1\lambda_2|v|\alpha\lambda_3\rangle
\tilde{X}^\mu_{\lambda_1\lambda_2:\lambda_3}
\nonumber \\
&=&0,
\label{stddm1n2}
\end{eqnarray}
where $\tilde{X}^\mu_{\alpha\beta:\gamma}=\langle 0|a^+_\alpha a^+_\beta a_\gamma|\mu\rangle$. The equation for 
$\tilde{X}^\mu_{\alpha\beta:\gamma}$ is given as
\begin{eqnarray}
\sum_\lambda((\langle\lambda|t|\alpha\rangle&-&\delta_{\alpha\lambda}\omega_\mu)\tilde{X}^\mu_{\lambda\beta:\gamma}
\nonumber \\
&+&\langle\lambda|t|\beta\rangle\tilde{X}^\mu_{\alpha\lambda:\gamma}
-\langle\gamma|t|\lambda\rangle\tilde{X}^\mu_{\alpha\beta:\lambda})
\nonumber \\
&+&\sum_{\lambda_1\lambda_2}
\langle\lambda_1\lambda_2|v|\alpha\beta\rangle\tilde{X}^\mu_{\lambda_1\lambda_2:\gamma}
\nonumber \\
&=&0.
\label{stddm2n2}
\end{eqnarray}
Since there are no higher-level transition amplitudes in an $A=2$ system, these two equations are also exact.
From Eq. (\ref{stddm1n2}) we obtain
\begin{eqnarray}
\frac{1}{2}\sum_{\mu\alpha\alpha'} (\langle\alpha'|t|\alpha\rangle
 &+&\delta_{\alpha\alpha'}\omega_\mu)x^\mu_{\alpha'}(x^\mu_\alpha)^*
 \nonumber \\
 &=&\sum_{\alpha\alpha'} \langle\alpha'|t|\alpha\rangle
n_{\alpha\alpha'}
\nonumber \\
&+&\frac{1}{2}\sum_{\lambda_1\lambda_2\lambda_3\lambda_4}\langle\lambda_1\lambda_2|v|\lambda_3\lambda_4\rangle
\rho_{\lambda_3\lambda_4\lambda_1\lambda_2}
\nonumber \\
&=&\langle 0|H|0\rangle,
\label{gsenergy}
\end{eqnarray}
where $n_{\alpha\alpha'}$ and $\rho_{\alpha\beta\alpha'\beta'}$ are exactly given by
\begin{eqnarray}
n_{\alpha\alpha'}&=&\sum_{\mu} x^\mu_{\alpha'}(x^\mu_\alpha)^*,
\label{n1x1}
\end{eqnarray}
\begin{eqnarray}
\rho_{\alpha\beta\alpha'\beta'}&=&\sum_{\mu}\tilde{X}^\mu_{\alpha'\beta':\beta}(x^\mu_{\alpha})^*.
\label{n2x2}
\end{eqnarray}
Equation (\ref{gsenergy}) corresponds to the relation between the total ground-state energy and the single-particle Green's function \cite{Fetter}.

\section{Applications to $^{16}$O}
\subsection{Calculational details}
In this paper, we make a first schematic application of our theory to proton hole states in $^{15}$N 
and proton particle states in $^{17}$F. 
We do not consider the corresponding neutron states because there are less 
experimental data.
We consider 
the $1s_{1/2}$, $1p_{3/2}$, $1p_{1/2}$, $1d_{5/2}$, $ 2s_{1/2}$, $1d_{3/2}$, $2p_{3/2}$, $2p_{1/2}$
$1f_{7/2}$ and $1f_{5/2}$ states for both protons and neutrons. The continuum states are discretized by confining the 
single-particle wavefunctions in a sphere of radius 12 fm. We use a simplified
residual interaction which consists only of the $t_0$ and $t_3$ terms of the Skyrme III force. Its strength is reduced by 20\% 
to put the spurious c.o.m motion of $^{16}$O at approximately zero energy in the standard RPA.
For the ground-state calculation of $^{16}$O in TDDM, we only use the bound single-particle
states, the $1s_{1/2}$, $1p_{3/2}$, $1p_{1/2}$, $1d_{5/2}$ and $ 2s_{1/2}$ states, and consider only the two particle - two hole
type correlations in $C_{\alpha\beta\alpha'\beta'}$. We also neglect the off-diagonal elements of $n_{\alpha\alpha'}$ between the $1s_{1/2}$ and $2s_{1/2}$
states. 
The $y_p^\mu$ amplitudes for the proton $2s_{1/2}$, $2p_{1/2}$ and $2p_{1/2}$ states are neglected
because their contributions are negligible.

\subsection{Ground state}
\begin{table}
\caption{Single-particle energies $\epsilon_\alpha$ and occupation probabilities 
$n_{\alpha\alpha}$ calculated in TDDM. The single-particle energies in HF are given in 
parentheses.}
\begin{center}
\begin{tabular}{c|ll|rr} \hline
 &\multicolumn{2}{c|}{$\epsilon_\alpha$ [MeV]}&\multicolumn{2}{c}{$n_{\alpha\alpha}$}\\ \hline 
orbit & proton & neutron  & proton & neutron  \\ \hline
$1s_{1/2}$ & $-32.5$ $(-32.1)$ & $-36.2$ $(-35.9)$ & 0.98 & 0.98  \\
$1p_{3/2}$ & $-18.3$ $(-18.2)$ & $-21.8$ $(-21.8)$ & 0.93 & 0.93  \\
$1p_{1/2}$ & $-12.3$ $(-12.0)$ & $-15.8$ $(-15.6)$ & 0.91 & 0.91  \\
$1d_{5/2}$ & $-3.8$ $(-3.8)$ & $-7.1$ $(-7.2)$ & 0.08 & 0.08   \\ 
$2s_{1/2}$ & $~~~1.2$ $(1.5)$ & $-1.5$ $(-1.2)$ & 0.02 & 0.02  \\ \hline
\end{tabular}
\label{tab1}
\end{center}
\end{table}

The occupation probabilities calculated in TDDM are shown in Table \ref{tab1}.
The largest deviation from the HF values ($n^0_{\alpha\alpha}$=1 or 0) is about 10\%,
which means that the ground state of $^{16}$O is a strongly correlated state.
A recent shell-model calculation by Utsuno and Chiba \cite{utsuno} also gives a similar 
result for the ground state of $^{16}$O.
The correlation energy $E_c$ in the ground state, which is
defined by $E_c=\sum_{\alpha\beta\alpha'\beta'}\langle\alpha\beta|v|\alpha'\beta'\rangle C_{\alpha'\beta'\alpha\beta}/2$, is $-19.6$ MeV.
A large portion of the correlation energy is compensated by the increase in the mean-field energy due to the fractional occupation
of the single-particle states. The resulting energy gain due to the ground-state correlations, which is given
by the total energy difference between HF and TDDM, is with  5.2 MeV relatively small. Such kind of scenario is similar to the one well known from BCS 
theory \cite{RS}.

\subsection{Spectral functions}
\begin{figure} 
\begin{center} 
\includegraphics[height=6cm]{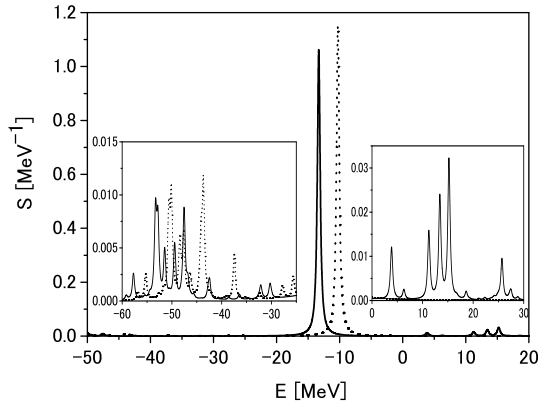}
\end{center}
\caption{Spectral function of 
the proton $1p_{1/2}$ state in $^{15}$N calculated in EoRPA (solid line). 
The dotted line shows the result in TDA.
The distributions are smoothed with an artificial width $\Gamma=0.5$ MeV.
The strength distribution in the positive energy region is due to the coupling to the backward amplitude $Y^\mu_{pp':h}$
and indicates the states in $^{17}$F. The small strength distributions in the positive and negative energy regions are shown
in the insets.} 
\label{p1} 
\end{figure} 
\begin{figure} 
\begin{center} 
\includegraphics[height=6cm]{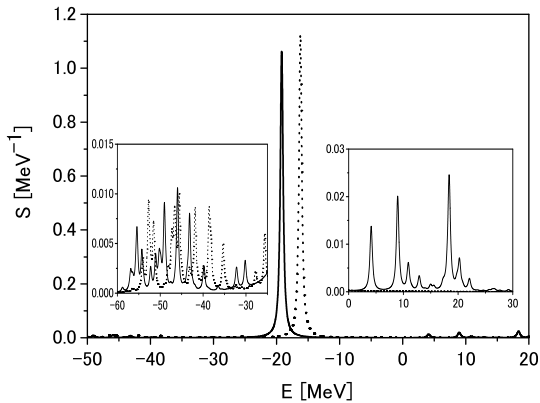}
\end{center}
\caption{Same as Fig. \ref{p1} but for the proton $1p_{3/2}$ state.} 
\label{p3} 
\end{figure}
In Figs. \ref{p1} and \ref{p3} the spectral functions of the proton $1p_{1/2}$ 
and $1p_{3/2}$ hole states in $^{15}$N calculated in EoRPA (Eq. (\ref{hrpa1})) (solid line) are shown, 
respectively, and compared with
the results in TDA (dotted line).
Since the results in oRPA are similar to the EoRPA results, they 
are not shown.
As already mentioned, in the EoRPA calculations we consider only the same $Y_{\alpha\beta:\gamma}^\mu$ amplitudes as 
those used in oRPA. 
To facilitate a comparison of various calculations, we smooth the distributions using
an artificial width $\Gamma=0.5$ MeV. As shown in Table \ref{tab1}
the HF energies of the proton $1p_{1/2}$ and $1p_{3/2}$ states are $-12.0$ MeV and $-18.2$ MeV, respectively.
In TDA the main peak is shifted upwards from the HF position due to the coupling to the configurations $Y_{hh':p}^\mu$
whose unperturbed energies are distributed below $-40$ MeV. In EoRPA (and oRPA) 
the main peak is slightly shifted downwards from the HF position due to the coupling to both $Y_{hh':p}^\mu$ and 
the backward amplitudes $Y_{pp':h}^\mu$ whose
unperturbed energies are located above 0 MeV (see Fig. \ref{p1} and 
Fig. \ref{p3}). 
The strength distribution in the positive energy region corresponds to
the states in $^{17}$F. The strengths of the main peak of the proton $1p_{1/2}$
state calculated in EoRPA, oRPA and TDA are 0.88, 0.82 and 0.95, respectively.
When the forward amplitude $Y_{hh':p}^\mu$ is neglected in oRPA,
the main peak is further shifted down to $-14.9$ MeV and has strength 0.89. This indicates that the coupling to the backward amplitudes
$Y_{pp':h}^\mu$ plays an important role in depleting the single-particle strength.
The effects of the ground-state correlations included in EoRPA play 
a role in slightly reducing the correlations in oRPA due to fractional occupation of the single-particle states. 
The sum of the strength 
of the proton $1p_{1/2}$ state distributed in the negative energy region
is 0.91 in EoRPA, which corresponds to $n_{\alpha\alpha}=0.91$ in TDDM, 
(see Table \ref{tab1}).
Thus the relation $n_{\alpha\alpha}=\sum_{\mu}|\langle\mu|a_\alpha|0\rangle|^2$ 
holds to a good approximation.
The single-particle strengths  of the main peak of the proton $1p_{3/2}$
state calculated in EoRPA, RPA and TDA are 0.88, 0.82 and 0.93, respectively.
The sum of the occupation probabilities 
of the proton $1p_{3/2}$ state distributed in the negative energy region
is 0.93 in EoRPA, which corresponds to $n_{\alpha\alpha}=0.93$ in TDDM. Summing the whole spectral weights in negative and positive energy regions gives, of course, the sum rule value of one.

\begin{figure} 
\begin{center} 
\includegraphics[height=6cm]{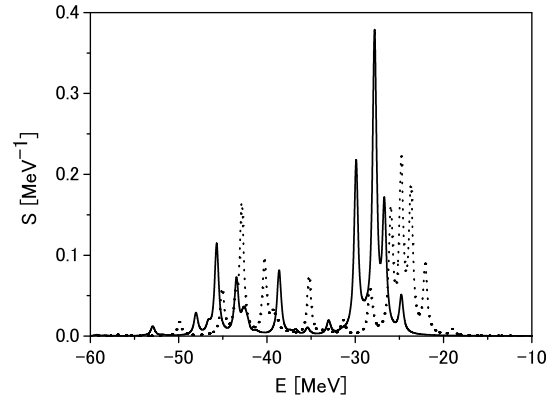}
\end{center}
\caption{Same as Fig. \ref{p1} but for the proton $1s_{1/2}$ state.} 
\label{s1} 
\end{figure}

The results for the proton $1s_{1/2}$ state are shown in Fig. \ref{s1}. The HF energy of the proton $1s_{1/2}$ hole state is $-32.1$ MeV.
The strength is fragmented due to the coupling to 
the configurations $Y_{hh':p}^\mu$: the unperturbed energy of the configuration 
$(1p_{1/2})^{-1}(1p_{3/2})^{-1}1d_{5/2}$ is about $-26~$MeV.
 Since the backward configurations $Y_{pp':h}^\mu$ are energetically well separated,
there is no significant difference between the TDA and oRPA results. Therefore, the oRPA result
is not shown in Fig. \ref{s1}. Comparing with the results obtained from Eq. (\ref{hrpa2}), 
we found that the $D_2$ term in the matrix $D$, which is given by $e\times N_{32}$ and describes the 
self-energy contributions to 
the one particle - two hole configurations, play a role in shifting the strength to lower energy region.
The summed occupation probability 
of the proton $1s_{1/2}$ state distributed in the negative energy region
is 0.98 in EoRPA, which corresponds to $n_{\alpha\alpha}=0.98$ in TDDM.

\begin{figure} 
\begin{center} 
\includegraphics[height=6cm]{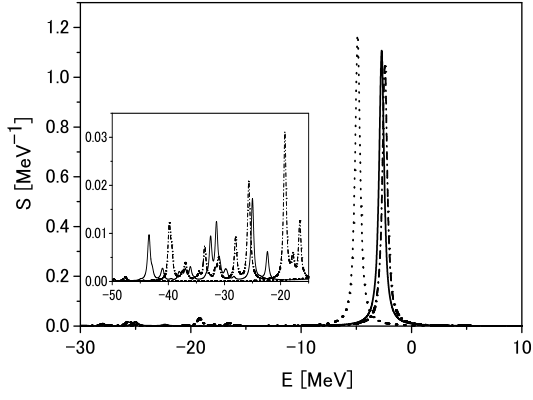}
\end{center}
\caption{Same as Fig. \ref{p1} but for the proton $1d_{5/2}$ state in $^{17}$F.
The dot-dashed line shows the result in oRPA.
The strength distribution below $-15$ MeV is due to the coupling to the backward amplitude $Y^\mu_{hh':p}$
and shows the states in $^{15}$N.} 
\label{d5} 
\end{figure}

The spectral function of the proton $1d_{5/2}$ state in $^{17}$F is shown in Fig. \ref{d5}.
The HF energy of the $1d_{5/2}$ state is $-3.8$ MeV.
The main peak is shifted downwards from the HF position in TDA due to the coupling to $Y_{pp':h}^\mu$, 
while, on the contrary, it is shifted upward in oRPA due to the additional coupling to the backward amplitudes $Y_{hh':p}^\mu$.
The ground-state correlations included in EoRPA play a role in slightly 
reducing correlations in oRPA.
The states located below the single-particle energy of the proton $1p_{1/2}$ state correspond 
to the states in $^{15}$N.
The summed occupation probability 
of the proton $1d_{5/2}$ state distributed below the proton $1p_{1/2}$ state
is 0.06 in EoRPA, while the corresponding value for $n_{\alpha\alpha}$ in TDDM is 
0.08.

\begin{figure} 
\begin{center} 
\includegraphics[height=6cm]{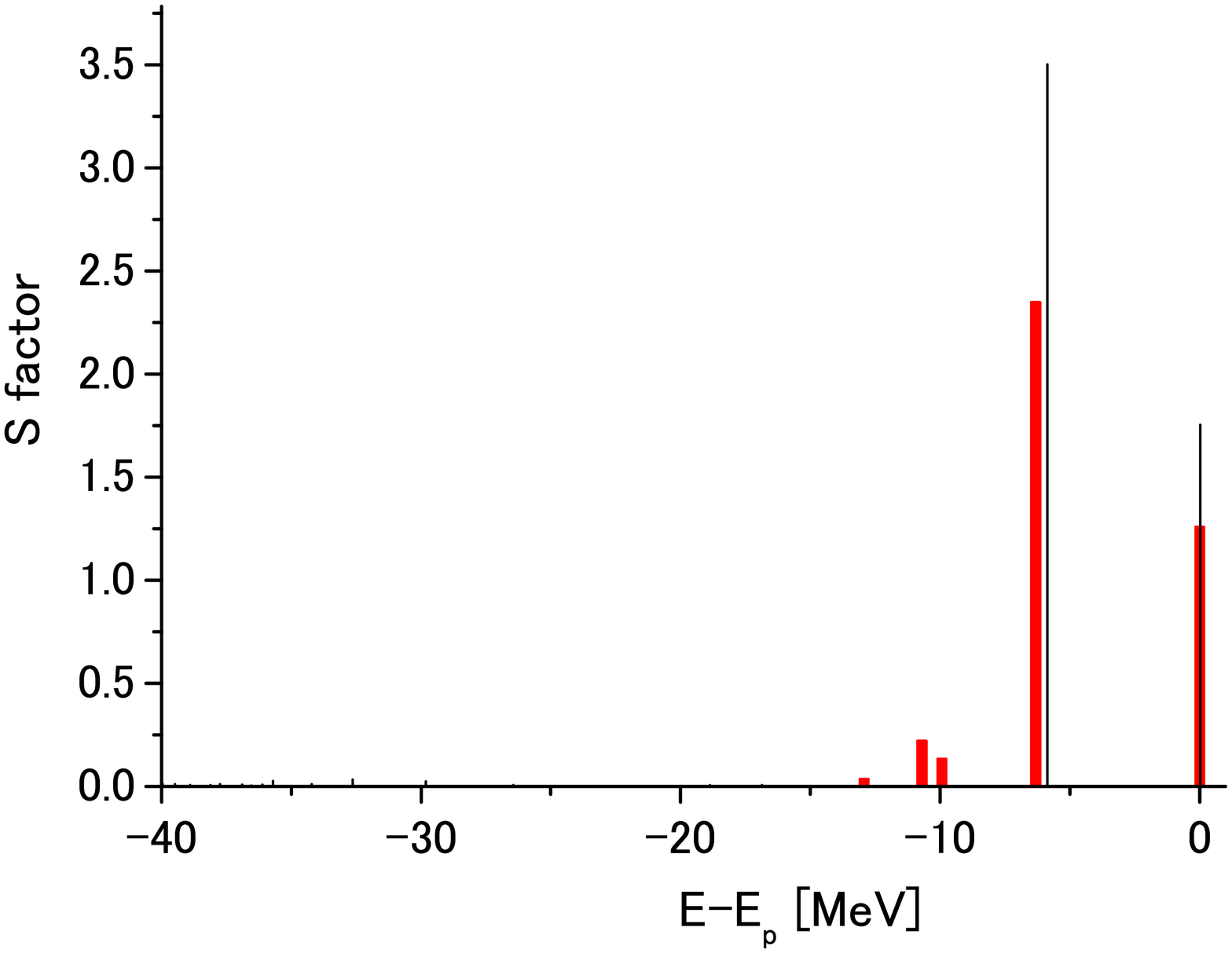}
\end{center}
\caption{Spectroscopic factors for the proton $1p_{1/2}$ and $1p_{3/2}$ states calculated 
in EoRPA are compared with experiment \cite{leus94} (red bars).} 
\label{pexp} 
\end{figure}
\begin{figure} 
\begin{center} 
\includegraphics[height=6cm]{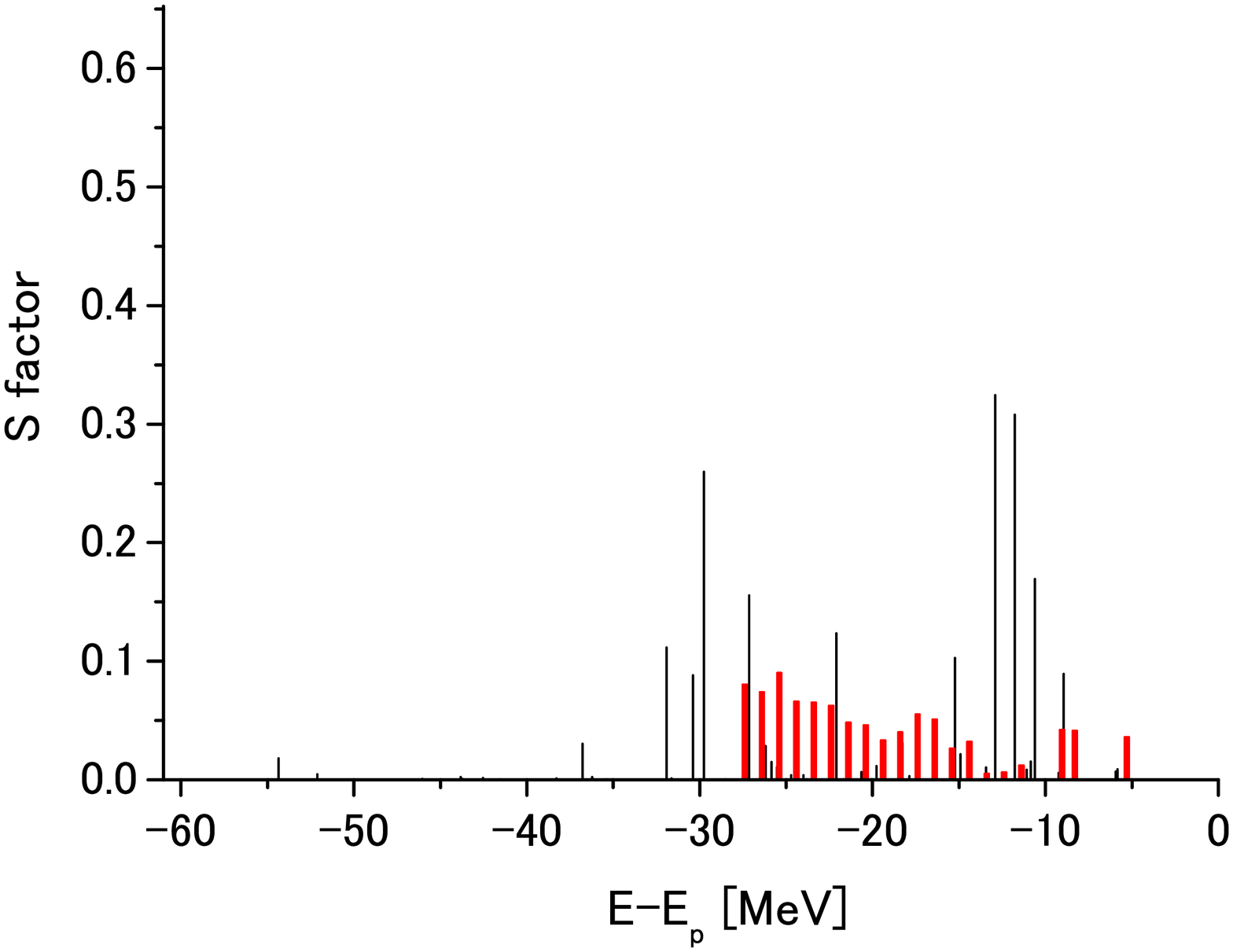}
\end{center}
\caption{Spectroscopic factors for the proton $1s_{1/2}$ and $2s_{1/2}$ states calculated 
in oRPA are compared with experiment \cite{leus94} (red bars).} 
\label{sexprpa} 
\end{figure}
\begin{figure} 
\begin{center} 
\includegraphics[height=6cm]{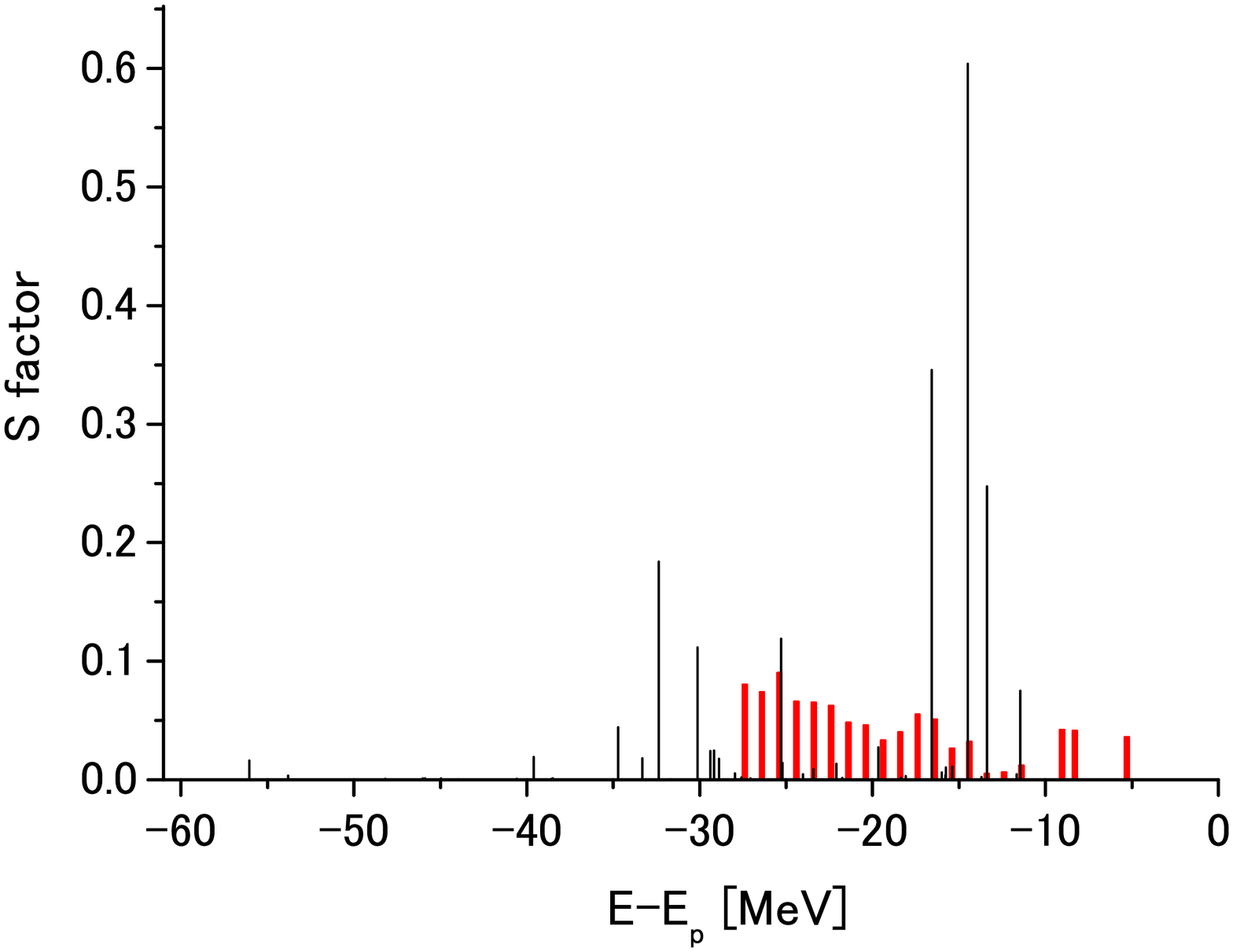}
\end{center}
\caption{Same as Fig. \ref{sexprpa} but for 
EoRPA.} 
\label{sexp} 
\end{figure}

\begin{figure} 
\begin{center} 
\includegraphics[height=6cm]{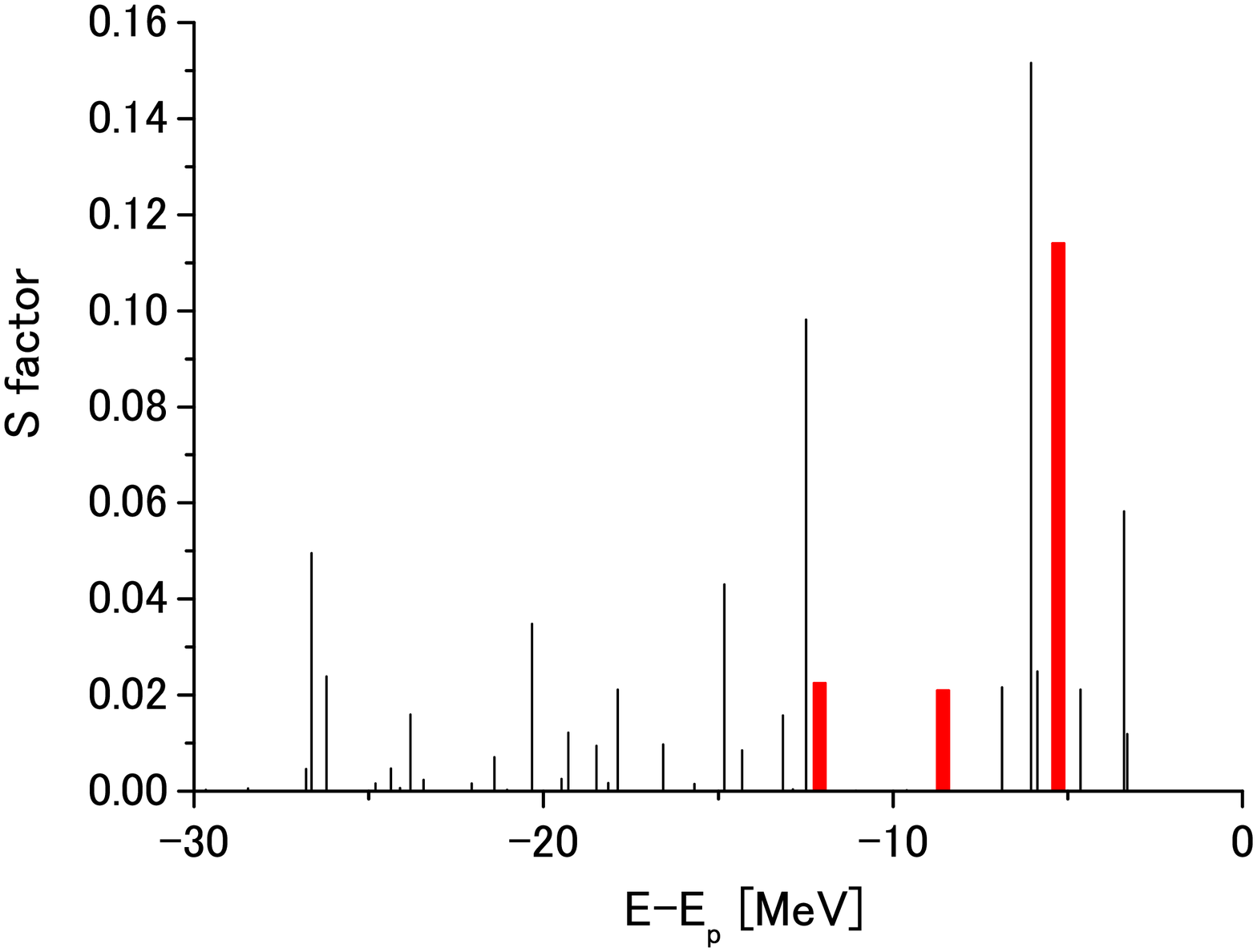}
\end{center}
\caption{Spectroscopic factors for the proton $1d_{5/2}$ and $1d_{3/2}$ states calculated 
in oRPA are compared with experiment \cite{leus94} (red bars).} 
\label{dexprpa} 
\end{figure}
\begin{figure} 
\begin{center} 
\includegraphics[height=6cm]{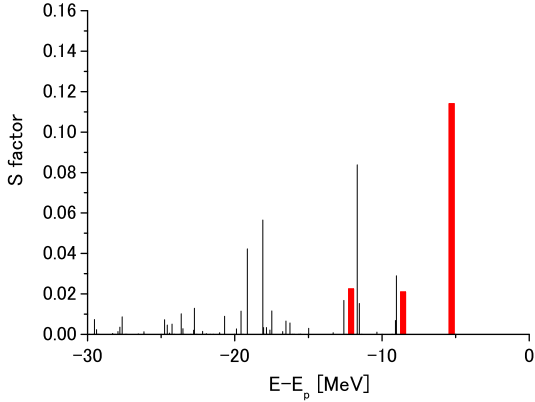}
\end{center}
\caption{Same as Fig. \ref{dexprpa} but for
EoRPA.} 
\label{dexp} 
\end{figure}

\subsection{Comparison with experiment}
The spectroscopic factors (defined by $(2j+1)\times$transition strength) calculated 
in EoRPA for the proton $1p_{1/2}$ and $1p_{3/2}$ states are compared with experiment \cite{leus94} (red bars) in Fig. \ref{pexp}.
The main peak of the proton $1p_{1/2}$ state is considered as the ground state of $^{15}$N and the hole-state energy is measured
from this threshold in the following. The results in EoRPA are reasonable though they overestimate the experimental data and cannot reproduce the strength distribution
around $-10$ MeV. This is a common feature of TDA and RPA-type calculations \cite{hrpa96,dick02}.
Studies on the effect of short-range correlations have predicted a strength reduction of about 
10\% in $^{16}$O \cite{muth,radi,fab}.
The spectroscopic factors for the proton $1s_{1/2}$ and $2s_{1/2}$ states calculated 
in oRPA is compared with experiment \cite{leus94} (red bars) in Fig. \ref{sexprpa}.
The EoRPA results are also compared with experiment (red bars) in Fig. \ref{sexp}.
Since the inclusion of ground-state correlations causes a downward shift of the strength, the 
agreement with the data becomes somewhat worse in EoRPA.
The strong fragmentation below $-15$ MeV cannot be reproduced in these oRPA 
and EoRPA calculations. Probably higher configurations are needed.
The spectroscopic factors for the proton $1d_{5/2}$ and $1d_{3/2}$ states calculated 
in oRPA are compared with experiment \cite{leus94} (red bars) in Fig. \ref{dexprpa}.
The EoRPA results are also compared with experiment in Fig. \ref{dexp}. 
Due to a downward shift of the strength, the agreement with the data is 
worsened in EoRPA. 
We point out that there is a similar situation in the first $3^-$ state in $^{16}$O.
The effects of the ground-state correlations can be included into the standard RPA  
using $n_{\alpha\alpha'}$ and $C_{\alpha\beta\alpha'\beta'}$ as in SCRPA \cite{Janssen}. The first $3^-$ state of $^{16}$O
calculated in this modified RPA scheme comes about 5 MeV higher than the result in the standard RPA. This is the
same situation as the EoRPA results shown above. We found that inclusion of the coupling of the particle-hole
amplitude to higher two-particle two-hole amplitudes brings down the first $3^-$ state to the right position \cite{prep}.
Therefore, 
more elaborate calculations using a larger number of the $Y_{\alpha\beta:\gamma}^\mu$ amplitudes and 
also the higher amplitudes $Y^\mu_{\alpha\beta\gamma:\lambda\lambda'}$ could shift
the strength upward and bring a better agreement with the data. 
Globally, one may say that the agreement of spectroscopic factors with experiment is only marginally satisfactory indicating the need for inclusion of higher configurations.

\subsection{Effects of other amplitudes}

\begin{figure}  
\begin{center} 
\includegraphics[height=6cm]{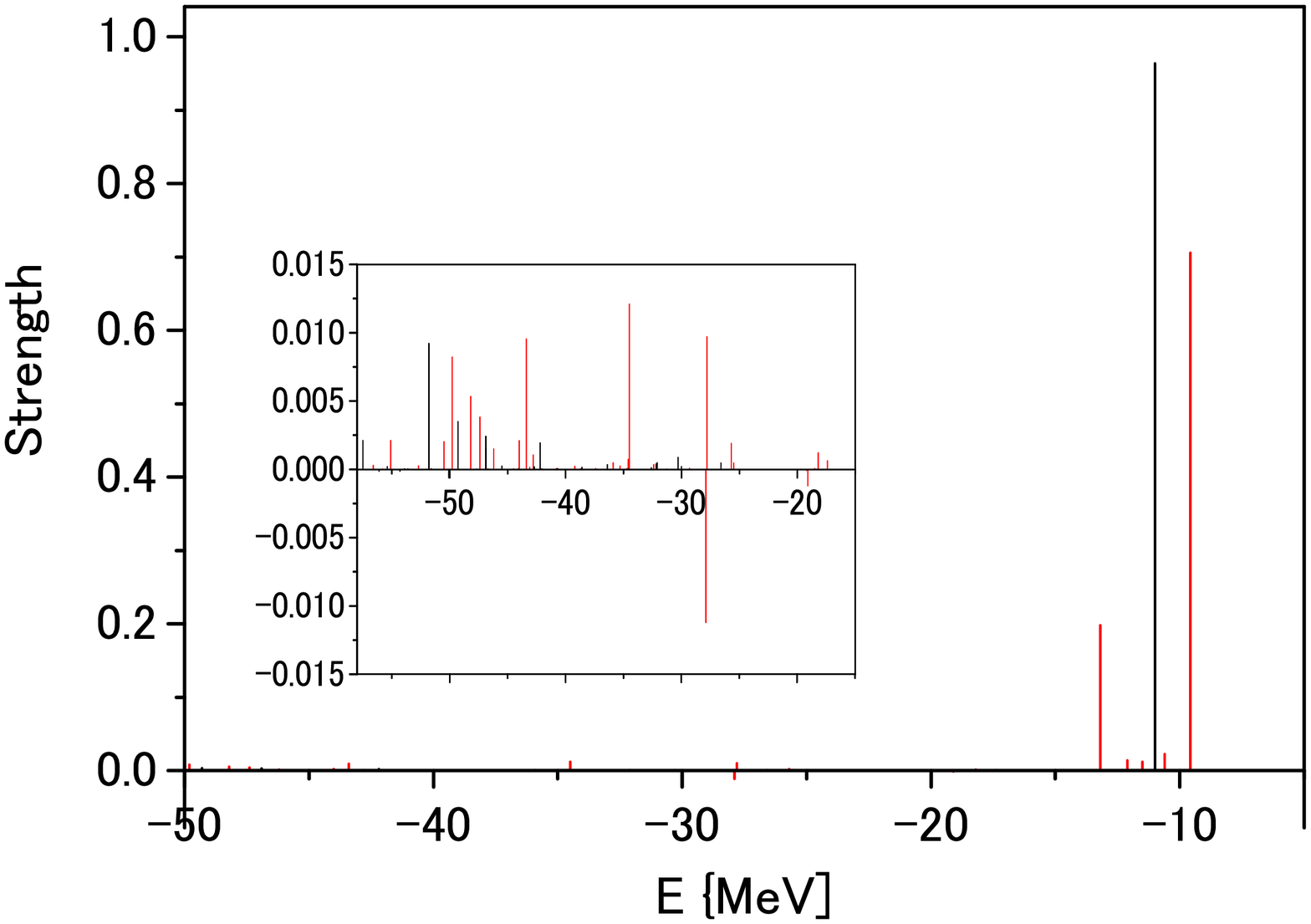}
\end{center}
\caption{Strength distribution of the proton $1p_{1/2}$ state calculated in EoRPA including the amplitude $Y^\mu_{hp:h'}$.
The red bars show the result of the calculation based on Eq. (\ref{hrpa}) 
where only the amplitudes $X^\mu_{hp:h'}$ and $X^\mu_{hh':p}$ are included under the assumption of the HF ground state.
Small strengths in the low-energy region are shown in the inset.} 
\label{erpap1} 
\end{figure}

\begin{figure} 
\begin{center} 
\includegraphics[height=6cm]{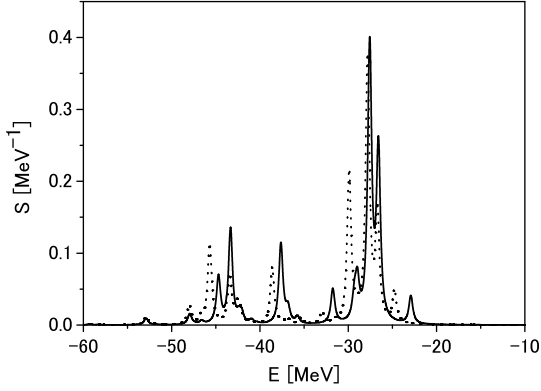}
\end{center}
\caption{Strength distribution of the proton $1s_{1/2}$ state calculated in EoRPA 
with (solid line) and without (dotted line) $Y^\mu_{hp:h'}$.} 
\label{erpas1} 
\end{figure}

We investigate the effects of inclusion of the amplitude $Y^\mu_{hp:h'}$ in EoRPA, which describes backward
scattering of a particle - hole pair in $Y^\mu_{hh':p}$.
We use for $Y^\mu_{hp:h'}$
the same truncated single-particle space as that used in the ground-state calculation
since it is important to include the self-energy contribution to all single-particle states in $Y^\mu_{hp:h'}$. 
To reduce the dimension size, 
we neglect the amplitude $Y^\mu_{pp':h}$.
The obtained result for the proton $1p_{1/2}$ state is shown in Fig. \ref{erpap1} and compared with the result (red bars)
of the calculation based on Eq. (\ref{hrpa}) where the ground state is assumed to be the HF ground state
and only the amplitudes $X^\mu_{hp:h'}$ and $X^\mu_{hh':p}$ are included. As shown in Fig. \ref{erpap1},
the inclusion of $X^\mu_{hp:h'}$ gives quite
unphysical results : the main peak is fragmented and some states have negative strength.
The reason for the fragmentation of the main peak is that 
unperturbed energies of some $X^\mu_{hp:h'}$ fall near the energy of the proton $1p_{1/2}$ state.
For example, the unperturbed energy of the configuration $(1s_{1/2})^{-1}2s_{1/2}(1p_{3/2})^{-1}$
that couples to the proton $1p_{1/2}$ state is $-12.4~$MeV, which is close to the energy of this state ($\epsilon_\alpha=-12.0$ MeV).
These unphysical properties are not seen in the EoRPA result. We consider that this is due both to the self-energy insertion to 
the configurations $Y^\mu_{hp:h'}$ and to their small normalization $N_{22}$. The energy of the configuration 
$Y^\mu_{hp:h'}$ is significantly shifted by the amount determined by the self energy and the normalization. This shift probably
plays a role in reducing the coupling to the single-hole state. 
We performed a similar EoRPA calculation for the proton $1s_{1/2}$ state, see Fig. \ref{erpas1} 
and the obtained result (solid line) is compared with the EoRPA
result  without $Y^\mu_{hp:h'}$  (dotted line). The coupling to $Y^\mu_{hp:h'}$ plays a role in shifting some strength upward,
which improves the agreement with the experiment. However,
we found that 
the inclusion of other amplitudes such as $Y^\mu_{hh':h''}$ and $Y^\mu_{hp:p'}$ brings unphysical fragmentation of the 
strength of the $1p_{1/2}$ state as seen in Fig. \ref{erpap1}.  
Therefore, it requires further investigation whether the amplitudes of $Y^\mu_{\alpha\beta:\gamma}$ with small normalizations 
should be included or not in EoRPA.

\begin{figure} 
\begin{center} 
\includegraphics[height=6cm]{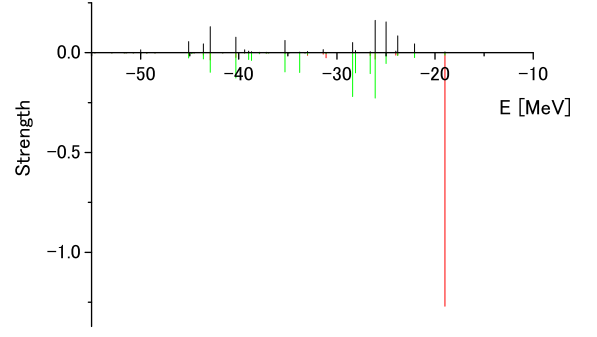}
\end{center}
\caption{Strength distribution of $|\langle \mu|P_za_\alpha|0\rangle| ^2$ in the states 
which couple to the proton $1s_{1/2}$ state. The upper part shows
the strength of the proton $1s_{1/2}$ state and the lower part $-|\langle \mu|P_za_\alpha|0\rangle| ^2$
(in arbitrary units)
for $\alpha=$ the proton $1p_{1/2}$ state (red bars) and the proton $1p_{3/2}$ state (green bars).
The strength $|\langle \mu|P_za_\alpha|0\rangle| ^2$ for $\alpha=1p_{3/2}$ is fragmented
due to the coupling to the configurations which have the particle-hole pairs with angular momentum $L=2~\hbar$. } 
\label{spurs1} 
\end{figure}
\begin{figure} 
\begin{center} 
\includegraphics[height=6cm]{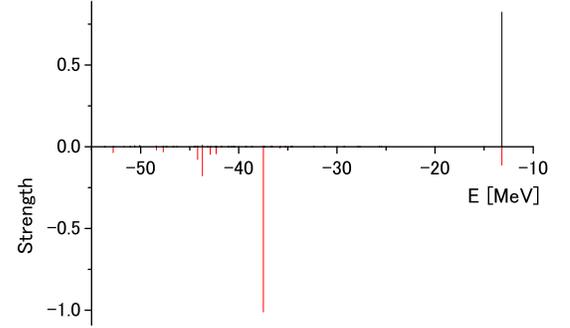}
\end{center}
\caption{Strength distribution of $|\langle \mu|P_za_\alpha|0\rangle |^2$ in the states 
which couple to the proton $1p_{1/2}$ state. The upper part shows
the strength of the proton $1p_{1/2}$ state and the lower part $-|\langle \mu|P_za_\alpha|0\rangle |^2$ (in arbitrary units)
for $\alpha=$ the proton $1s_{1/2}$ state (red bars).} 
\label{spurp1} 
\end{figure}
\begin{figure} 
\begin{center} 
\includegraphics[height=6cm]{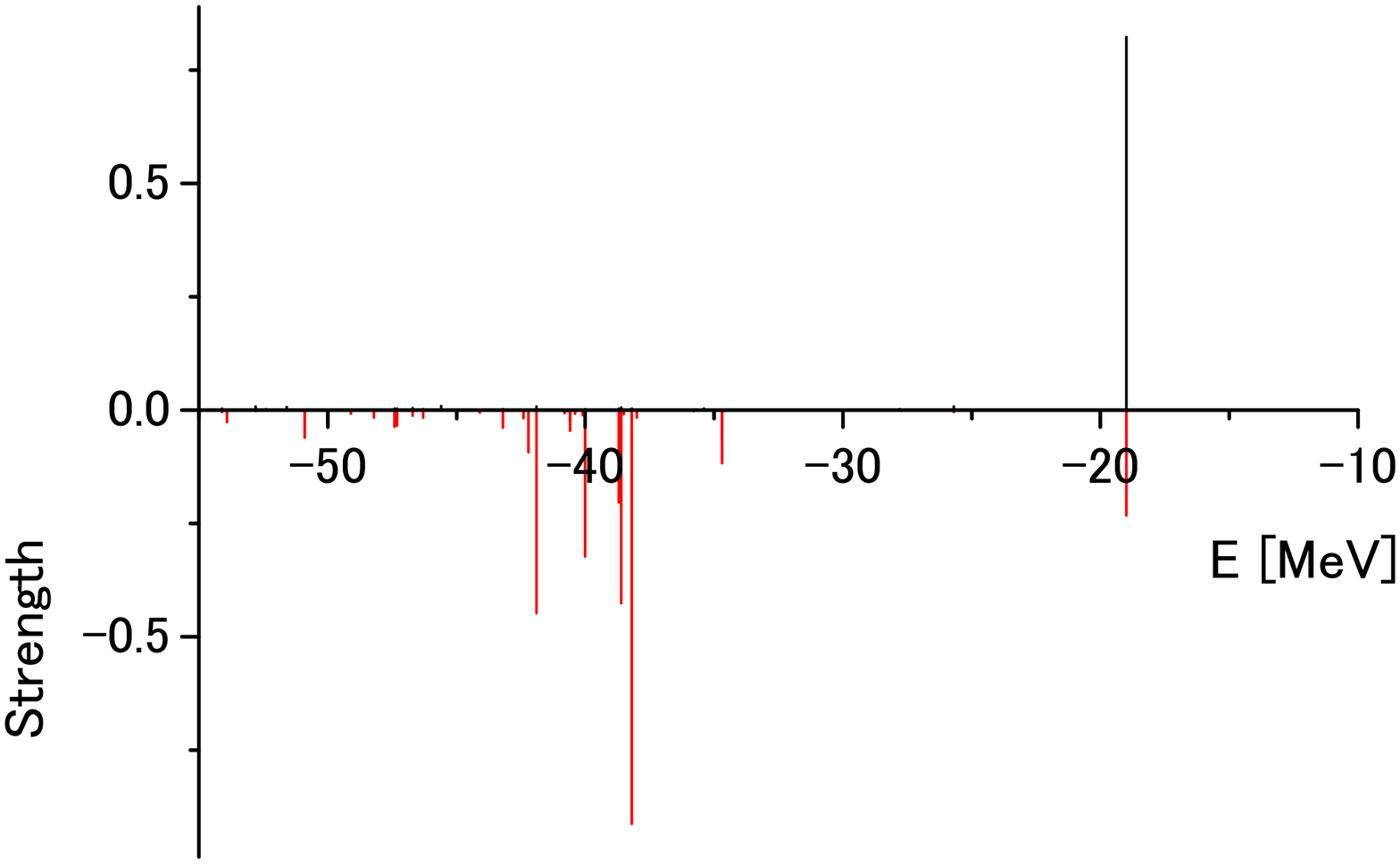}
\end{center}
\caption{Same as Fig. \ref{spurp1} but for the proton $1p_{3/2}$ state.} 
\label{spurp3} 
\end{figure}

\subsection{Center of mass motion of $^{16}$O}
Finally we discuss the coupling of a hole state to the c.o.m. motion of the core nucleus $^{16}$O using oRPA.
The strength distribution of $|\langle \mu|P_za_\alpha|0\rangle| ^2$ among the states 
which couple to the proton $1s_{1/2}$ state is shown in Fig. \ref{spurs1}, where the proton 
$1p_{1/2}$ and $1p_{3/2}$ states are taken for $\alpha$. The upper part of Fig. \ref{spurs1} shows
the strength of the proton $1s_{1/2}$ state (the same as Fig. \ref{sexprpa}) and the lower part $-|\langle \mu|P_za_\alpha|0\rangle| ^2$
for $\alpha=1p_{1/2}$ (red bars) and $1p_{3/2}$ (green bars).
The strength $|\langle \mu|P_za_\alpha|0\rangle| ^2$ for $\alpha=1p_{1/2}$ is concentrated in a single state and
the coupling of the proton $1s_{1/2}$ state to this state is negligible. 
Therefore, this state may be interpreted as a spurious mode
consisting of a pure c.o.m motion of $^{16}$O and the proton $1p_{1/2}$ state, though it 
is located about 7 MeV below the single-particle energy of the proton $1p_{1/2}$ state ($-12.0$ MeV).
This energy shift from the single-particle energy is related to the fact that the TDA calculation for
the c.o.m motion of $^{16}$O gives the excitation energy of 7.7 MeV. 
This fact may look perturbing, since we know that the spurious mode of the core comes at zero energy
in the standard RPA. 
As already mentioned above, it is not guaranteed in an odd system that the c.o.m motion of the core comes at zero energy, and 
the coupling of the spurious mode to the physical spectrum is very week, 
so that the position of the spurious mode is not so perturbing. 
We also performed a TDA calculation for the c.o.m motion of $^{208}$Pb using the single-particle states and the residual interaction
which put the spurious mode at zero energy in RPA and found that it comes at 4.7 MeV. This suggests that the c.o.m of very heavy systems
could come close to zero excitation energy even in TDA.
The strength $|\langle \mu|P_za_\alpha|0\rangle |^2$ for $\alpha=1p_{3/2}$ is rather fragmented.
Since the dominant components of the c.o.m motion of 
the core consists of the transitions from the $1p_{3/2}$ to $1d_{5/2}$ states,
the large fragmentation of $|\langle \mu|P_za_\alpha|0\rangle| ^2$ for $\alpha=1p_{3/2}$ 
is explained by the coupling to the configurations with different coupling schemes of angular momenta:
in the case of $\alpha=1p_{3/2}$, $Y^\mu_{\alpha h:p}$ consisting of the particle - hole pairs
$(1p_{3/2})^{-1}1d_{5/2}$ with angular momentum $L=1~\hbar$ can couple not only to $Y^\mu_{\alpha h:p}$ which has the same particle - hole
pairs with $L=2~\hbar$ but also to $Y^\mu_{\alpha h:p}$ consisting of the particle - hole pairs
$(1p_{1/2})^{-1}1d_{5/2}$ with $L=2~\hbar$. In the case of $\alpha=1p_{1/2}$ the particle - hole pairs $(1p_{3/2})^{-1}1d_{5/2}$ 
in $Y^\mu_{\alpha h:p}$ can have only $L=1~\hbar$ and does not couple to nearby configurations.
The exchange effect may also play a role in weakening the coherence of the c.o.m motion of $^{16}$O for $\alpha=1p_{3/2}$:
we tried an oRPA calculation for $\alpha=1p_{3/2}$ where all exchange terms are neglected and observed
the appearance of such a coherent state as that observed for $\alpha=1p_{1/2}$.
Thus, in the case of the proton $1s_{1/2}$ state and $\alpha=1p_{3/2}$
the c.o.m motion of the core nucleus is embedded in the physical states of the $A-1$ nucleus and cannot be neglected. 
The distributions of $|\langle \mu|P_za_\alpha|0\rangle |^2$ among the states which couple to
the proton $1p_{1/2}$ and $1p_{3/2}$ states are shown in Figs. \ref{spurp1} and \ref{spurp3}, respectively,
where $\alpha$ is the proton $1s_{1/2}$ state. 
In the case of the proton $1p_{1/2}$ state 
the strength $|\langle \mu|P_za_\alpha|0\rangle |^2$ is concentrated in a single state and the coupling of
the proton $1p_{1/2}$ state to this state is negligible. As in the case of the proton $1s_{1/2}$ state 
this state may be interpreted as a spurious mode
consisting of the c.o.m motion of $^{16}$O and the proton $1s_{1/2}$ state, though it 
is located about 5 MeV below the single-particle energy of the proton $1s_{1/2}$ state ($-32.1$ MeV). 
The fragmentation of $|\langle \mu|P_za_\alpha|0\rangle |^2$ for the proton $1p_{3/2}$ state is larger than that
for the proton $1p_{1/2}$ state. This is explained by the coupling to the configurations
with different angular momentum couplings: in the case of the $1p_{3/2}$ state 
the particle - hole pairs $(1p_{3/2})^{-1}1d_{5/2}$ in  $Y^\mu_{\alpha h:p}$
can carry angular momentum $L=1~\hbar$ and $2~\hbar$,
whereas the pairs cannot have $L=2~\hbar$ in the case of the $1p_{1/2}$ state.

\section{Summary}

In this paper, we took up the old subject of the RPA approach to odd particle 
systems. Those equations based on the usual equation of motion method (EoM) 
encountered in the past some difficulties \cite{hrpa96}. This gave raise to the so-
called 
Faddeev-RPA (FRPA) approach \cite{dick02}. However, whenever the RPA breaks down, so 
does FRPA. We located some of the difficulties of the old odd particle RPA (oRPA) and 
proposed some cure, limiting the configuration space to the normalizable 
subspace. We showed that p-RPA and h-RPA equations give identical results 
what is very similar to the property of pp(hh)RPA for even systems \cite{RS}. We 
also discussed the influence of the c.o.m. motion of the core on the odd 
particle (p or h). No difficulty with a break down seems to arise. It turned 
out that the recoil of the core influences the spectrum. 
This aspect may be most important for rotational states in deformed nuclei 
where the so-called spurious modes are, in fact, physical states. We also 
showed how to include ground state correlations explicitly in EoRPA, similar 
to what is done in TDDM, on top of the oRPA equations.
We made a first schematic application, using a simplified Skyrme force, to the 
hole and particle states around $^{16}$O. We compared Tamm Dancoff, oRPA, and 
EoRPA solutions. It was shown that in some cases all three approaches give 
very similar results but that in others the influence of extra RPA correlations 
were significant. The comparison with experiment is sufficiently encouraging 
to develop this kind of RPA approach further. In fact, the spirit of oRPA 
is quite close to second RPA. We encountered problems for the odd 
systems e.g. that the spectrum becomes too much shifted downwards. Such open 
problems may be a subject for the future. Also the connection between a common 
RPA vacuum in the even and odd systems, as proposed recently \cite{jemai}, 
may be 
an interesting further line of research.
\\
\\

\appendix
\section{}
\begin{eqnarray}
a(\alpha:\alpha')=\epsilon_\alpha\delta_{\alpha\alpha'}
\end{eqnarray}
\begin{eqnarray}
b(\alpha\beta\gamma:\alpha')&=&\sum_{\lambda}\langle\alpha'\lambda|v|\alpha\beta\rangle_An_{\gamma\lambda}
\nonumber \\
&-&\sum_{\lambda\lambda'}[\langle\alpha'\lambda'|v|\alpha\lambda\rangle_A
(n_{\lambda\beta}n_{\gamma\lambda'}+C_{\gamma\lambda\lambda'\beta})
\nonumber \\
&+&\langle\alpha'\lambda'|v|\lambda\beta\rangle_A
(n_{\lambda\alpha}n_{\gamma\lambda'}+C_{\gamma\lambda\lambda'\alpha})
\nonumber \\
-\langle\alpha'\gamma|&v&|\lambda\lambda'\rangle_A(n_{\lambda\alpha}n_{\lambda'\beta}
+\frac{1}{2}C_{\lambda\lambda'\alpha\beta})],
\end{eqnarray}
\begin{eqnarray}
c(\alpha:\alpha'\beta'\gamma')&=&\langle\alpha'\beta'|v|\alpha\gamma'\rangle
\end{eqnarray}
\begin{eqnarray}
d&(&\alpha\beta\gamma:\alpha'\beta'\gamma')=(\epsilon_\alpha+\epsilon_\beta-\epsilon_\gamma)
\delta_{\alpha\alpha'}\delta_{\beta\beta'}\delta_{\gamma\gamma'}
\nonumber \\
&+&\frac{1}{2}\langle\alpha'\beta'|v|\alpha\beta\rangle_A\delta_{\gamma\gamma'}
\nonumber \\
&+&\sum_{\lambda}[\langle\lambda\alpha'|v|\alpha\gamma'\rangle_An_{\gamma\lambda}\delta_{\beta\beta'}
%\nonumber \\
-\langle\lambda\alpha'|v|\beta\gamma'\rangle_An_{\gamma\lambda}\delta_{\alpha\beta'}
\nonumber \\
&+&\langle\gamma\beta'|v|\lambda\gamma'\rangle_An_{\lambda\alpha}\delta_{\beta\alpha'}
%\nonumber \\
-\langle\gamma\beta'|v|\lambda\gamma'\rangle_An_{\lambda\beta}\delta_{\alpha\alpha'}
\nonumber \\
&-&\frac{1}{2}\delta_{\gamma\gamma'}(\langle\alpha'\beta'|v|\alpha\lambda\rangle_An_{\lambda\beta}
%\nonumber \\
+\langle\alpha'\beta'|v|\lambda\beta\rangle_An_{\lambda\alpha})].
\end{eqnarray}

The norm matrix $N_{22}$ is given as 
\begin{eqnarray}
N_{22}(\alpha\beta\gamma&:&\alpha'\beta'\gamma')=
(\delta_{\alpha\alpha'}\delta_{\beta\beta'}-\delta_{\alpha\beta'}\delta_{\beta\alpha'})n_{\gamma'\gamma}
\nonumber \\
&+&\delta_{\gamma\gamma'}(n_{\alpha\alpha'}n_{\beta\beta'}-n_{\alpha\beta'}n_{\beta\alpha'}
+C_{\alpha\beta\alpha'\beta'})
\nonumber \\
&-&\delta_{\alpha\alpha'}(n_{\gamma'\gamma}n_{\beta\beta'}+C_{\gamma'\beta\gamma\beta'})
\nonumber \\
&-&\delta_{\beta\beta'}(n_{\gamma'\gamma}n_{\alpha\alpha'}+C_{\gamma'\alpha\gamma\alpha'})
\nonumber \\
&+&\delta_{\alpha\beta'}(n_{\gamma'\gamma}n_{\beta\alpha'}+C_{\gamma'\beta\gamma\alpha'})
\nonumber \\
&+&\delta_{\beta\alpha'}(n_{\gamma'\gamma}n_{\alpha\beta'}+C_{\gamma'\alpha\gamma\beta'}).
\end{eqnarray}

\end{document}